\documentclass[12pt,pra,aps,amssymb,amsfonts,amsmath,tightenlines]{revtex4}
\usepackage{dsfont}
\usepackage{graphicx}
\usepackage{amssymb,amsfonts,amsthm}
\usepackage{color}
\usepackage{verbatim}
\usepackage[normalem]{ulem}
\usepackage{enumerate}
\usepackage{mathrsfs}
\usepackage{bm,float}
\usepackage{mathtools}
\usepackage{extarrows}

\usepackage{textgreek}
\usepackage{lgreek}

\newtheorem{Proposition}{Proposition}
\newtheorem{Theorem}{Theorem}

\newtheorem{Definition}{Definition}

\usepackage{nicefrac}

\newcommand{\ri}{\mbox{$\rm i$}}

\usepackage{multirow}
\usepackage{rotating}
\usepackage{caption}
\begin{document}

\title{Mathematical Foundations of Complex Tonality}
\author{Jeffrey~R.~Boland$^1$ and Lane~P.~Hughston$^{2}$}

\affiliation{$^1$Syndikat LLC, 215 South Santa Fe Avenue, Los Angeles, California 90012, USA\\
$^{2}$Department of Computing, Goldsmiths University of London, New Cross, London SE14 6NW, UK\\}

\begin{abstract}
\noindent 
Equal temperament, in which semitones are tuned in the  irrational  ratio of $2^{1/12} : 1$, is best seen as a serviceable compromise, sacrificing purity for flexibility. Just intonation, in which intervals are given by products of powers of $2$, $3$, and $5$, is more natural, but of limited flexibility. We propose a new scheme in which ratios of Gaussian integers form the basis of an abstract tonal system. The tritone, so problematic in just temperament, 
given ambiguously by the ratios $\tfrac{45}{32}$, $\tfrac{64}{45}$, $\tfrac{36}{25}$, $\tfrac{25}{18}$, none satisfactory, 
is in our scheme represented by the complex ratio $1 + \rm{i} : 1$.   The major and minor whole tones, given by intervals of $\tfrac{9}{8}$ and $\tfrac{10}{9}$, can each be factorized into products of complex semitones, giving us a major complex semitone $\tfrac{3}{4}(1 + \rm{i})$ and a minor complex semitone $\tfrac{1}{3}(3 + \rm{i})$. 
The perfect third, given by the interval $\tfrac{5}{4}$, factorizes into the product of a complex whole tone $\tfrac{1}{2}(1 + 2\rm{i})$ and its complex conjugate. 
Augmented with these supplementary tones, the resulting scheme of complex intervals based on products of low powers of Gaussian primes leads to the construction of a complete system of
major and minor scales in all keys.
\vspace{-0.2cm}
\\
\begin{center}
\end{center}
\end{abstract}

\maketitle
%
\section{Introduction}
\noindent For centuries, musicians and mathematicians have struggled with the challenges of temperament and tonality. The problem is that there is a paradox at the heart of  Western music. The paradox is that no tuning system exists for keyboard instruments such that all the harmonious intervals are perfectly tuned. Here we propose a radical solution based on the notion of {\em complex tonality}. 
We begin with an overview of the just system of intonation, which originated in antiquity and was much studied in our era during the 16th, 17th and 18th centuries, reaching a high point in the work of Euler (1739, \!1774). We can define the just system as the totality of the various ways of representing tones by the vibrational frequencies of strings of various lengths, each such frequency being given by products of powers of the primes 2, 3, and 5 times some fixed frequency, which we call the tonal center.  Over time, many different ways of choosing the twelve notes of the chromatic scale in the just system have been proposed. Our goal is to construct a complex analogue of the just system, with broad applications, based on the ratios of certain Gaussian integers, in such a way that some of the  insufficiencies of the just system can be resolved. 

\newpage
Among the various diatonic scales that have been considered in the just system, the so-called {\it intense diatonic scale} of Claudius Ptolemeus is especially notable (Solomon 2000). Ptolemy's scale, expressed in units of time such that the frequency of middle $C$ is unity, takes the form
\begin{eqnarray} \label{just diatonic scale}
C = 1, \, \,
D = \tfrac{9}{8}, \, \,
E = \tfrac{5}{4}, \, \,
F = \tfrac{4}{3}, \, \,
G = \tfrac{3}{2}, \, \,
A = \tfrac{5}{3}, \, \,
B = \tfrac{15}{8}, \, \,
C' = 2.
\end{eqnarray} 
One can then ask how the Ptolemaic scale can be embedded in a chromatic scale. There are many possibilities, but one good example is the following, due to Johannes Kepler, which is known  as Kepler's Monochord No.~2 (Barbour 1953), here transposed down a fifth: 
\begin{eqnarray} \label{flat scale}
&&
C = 1, \, \,
D_{\flat} = \tfrac{16}{15}, \, \,
D = \tfrac{9}{8}, \, \,
E_ {\flat}= \tfrac{6}{5}, \, \,
E = \tfrac{5}{4},  \, \,
F = \tfrac{4}{3},  \, \,
G_ {\flat} = \tfrac{45}{32},  \, \,
\nonumber \\ && \quad \, \, G = \tfrac{3}{2},  \, \,
 A_ {\flat} = \tfrac{8}{5},  \, \,
A = \tfrac{5}{3},  \, \,
B_ {\flat} = \tfrac{16}{9},  \, \,
B = \tfrac{15}{8},  \, \,
C' = 2.
\end{eqnarray} 
In Section II we look at the just system in some detail and we sketch the arguments leading to Ptolemy's diatonic scale and to Kepler's chromatic scale. Among the defects of the just system, clearly visible in Kepler's scale, are the multiplicities of primes required for the tritone interval from $C$ to $F_{\sharp}$. Also troubling, when one looks further, are the multiple tunings for the notes $D$ and $B_{\flat}$, and the unpleasant ``wolf\," tunings of certain intervals. 

As a first step forward, we propose an alternative approach to building up scales in the just system by use of an analogy with chemistry. In Section III, {\em Atoms, molecules and particles}, we generate the just scale by the aggregation of a small number of whole tone and semitone intervals, which we call atoms.  We show that one recovers the seven modal scales (in their modern form) by permutations of the aggregation sequence of the given whole tones and semitones. This leads to a consistent assignment of all the tones of the chromatic scale with the exception of $F_{\sharp}$. The equal temperament system, which evolved as a practical response to the difficulties of the just system, can also be cast into an ``atomic" framework by use of the chemical analogy, but one loses the mathematically pleasant representation of harmonious intervals as products of powers of primes. One is thus motivated to delve more deeply back into the just system, with a view to generalizing it in some way.  

In Section IV, {\em Chromatic tone systems}, we turn to the idea of constructing the just chromatic scale by the aggregation of three types of semitones. In Theorem 1, we use the aggregative approach to give a complete description of the various possible three-semitone chromatic extensions of Ptolemy's scale. More precisely, we ask for all systems of three rational semitones with the property that the chromatic scales assembled by aggregating such tones hit all seven values of the Ptolemaic scale. We prove the existence of exactly three distinct families of solutions, each in thirty-two variations, giving ninety-six scales in all. There is an overall integrity to the three families thus arising, since the semitones in the various systems exchange syntonic commas in forming whole tones, in analogy with the behavior of electrons in atomic physics. We present some applications of the resulting classification scheme to several of the known historical just schemes catalogued by Barbour (1953). Nonetheless, one is forced to admit that the three triples of rational semitones underlying this scheme contain high multiplicities of primes, hence contributing to the well-known perceived difficulties in the sonorities of the just system.

Following our description of the various just extensions of Ptolemy's scale using three rational semitones, and taking note of its shortcomings, but at the same time taking advantage of its methods, we introduce the idea of {\em complex tonality}, in which tones take on complex values given by ratios of Gaussian integers. 

The complex interval systems thereby arising can be regarded as the basis for a new abstract form of music, but can also be used as a new way of analyzing the harmonic and melodic structures of the established corpus of Western music. Our construction proposes that the tritone interval from $C$ to $F_{\sharp}$ can be given the value $1 + \ri$, a single power of a single complex prime, which can thus be seen as  the most consonant interval in the system.

In Section V, {\em Complex diatonic scales}, we pursue an analogy with our exposition of the just system in terms of atoms and molecules, and exhibit a set of complex whole tones and semitones, seven in total, including the $X$, $Y$, and $Z$ tones of the real just system, all involving small multiplicities of complex primes, sufficient to enable us to write down aggregation sequences for the complex analogues of the twelve major diatonic scales.

In Section VI, {\em Well-tempered complexity}, we turn to the mathematical foundations required for further exploration of the complex system. We summarize various properties of the Gaussian integers and in Definition 1 we introduce the notion of a complex $n$-limit tone, one that is a product of powers of Gaussian integers that are prime factors of rational integers of magnitude no more than $n$. This is a new idea that generalizes the existing theory of real $n$-limit tones very naturally. 
Specializing to the case of complex $5$-limit tones, we work out the general form of a complex $5$-limit phase factor in Proposition 1 and we point out that the set of complex $5$-limit intervals modulo units and complex $5$-limit phase factors forms a Generalized Interval System in the sense of Lewin (1987). 

In Section VII, {\em Complex chromatic scales}, we show that a complex chromatic scale incorporating all the notes of the twelve given complex diatonic scales can be constructed by aggregation of three complex semitones. This construction is carried out  in such a way that the chromatic scale extends the harmonious values of Ptolemy's diatonic scale, leading to our main result, Theorem 2, which gives a full description of the totality of complex systems that (a) embrace the harmonious Ptolemaic values, (b) admit three complex semitones as generators, and (c) observe a certain  symmetry condition that ensures that the first six tones aggregate to the value $1 + \ri$ for $F_{\sharp}$. We establish the existence and uniqueness of three such complex systems, the first of which is of particular interest on account of its full use of the factorization of the number five and the abundance of intervals that it exhibits involving low multiplicities of primes.

In Section VIII, {\em Complex Pythagorean systems}, we consider the possibility of an analogous construction for two-semitone complex $3$-limit scales by fixing the values of the Pythagorean pentatonic scale and asking that the system  take the value $1+\ri$ at $F_{\sharp}$. The existence and uniqueness of such a scheme is shown in Theorem 3. In Section IX, {\em Back to Reality}, we return to the study of real tuning systems and ask for the construction of ``shadow" systems based on the magnitudes of the intervals arising in the complex schemes. This program can be carried out both for complex 3-semitone systems and for complex 2-semitone systems. The results are summarized in our Theorems 4 and 5. 

In Section X, {\em Applications, conjectures, and conclusions}, we wrap up and speculate on some applications to the theory of harmony. 

In Tables I, II, III, and IV we lay out the values of the tones of the chromatic scales for the systems that we have considered, alongside the associated aggregation sequences, showing the variations that arise in some cases.

In Tables V and VI we present a list of intervals occurring frequently in our analysis of complex tone systems. Our brief list can be compared to the more complete catalogues of real tones  provided by Ellis (1864), Helmholtz  (1885), Longuet-Higgins (1987), Halu$\check{\rm{s}}$ka (2004), and others, with which there is of course some overlap.
%
\section{Just Intonation}
\noindent  The question of which musical intervals are harmonious and which are dissonant is at least, in part, a matter of physiology, convention and culture---though it is undeniable that mathematical elements enter (albeit unconsciously) into musical aesthetics in an essential way, even at the level of tonality (Euler 1739, Helmholtz 1885, Longuet-Higgins 1987, Penrose 2000). The tradition in Western music is that the unison, the octave, the fifth, the fourth, the major third, the minor third, the major sixth and the minor sixth are harmonious. 
But the whole tone, the semitone, the major seventh, the minor seventh and the tritone are all to varying degrees dissonant. 

It has been known since ancient times that there is a relation between harmonious intervals and the ratios of the frequencies at which strings vibrate.
Suppose we fix a length of string so that it will vibrate at about 261.62 Hertz, so-called middle $C$ (or $C_4$) on the piano.  
If we halve the length of string then the frequency of vibration is doubled,  and we get $C_5$, the note which is one octave above middle $C$.
If we take one third of the length of string then the frequency is trebled, and we get $G_5$, the $G$ that lies a fifth above $C_5$. 

One fourth of the length of the string vibrates at four times the original frequency, and one fifth vibrates at five times the original. 
A string vibrating at a frequency four times $C_4$ sounds at $C_6$, two octaves above $C_4$, and a string vibrating at five times $C_4$ sounds at $E_6$.
It follows that the frequency of $G_4$ will be half that of $G_5$. 
If we use units of time such that the frequency of $C_4$ is unity, then 
$G_5 = 3$ and we deduce that $G_4 = \frac{3}{2}$. 
Likewise since  $E_6 = 5$ we deduce that the $E$ just above middle $C$ sounds at $E_4 = \frac{5}{4}$. 
We conclude that the interval of a fifth is given by the ratio $\frac{3}{2}$, the major third is the ratio $\frac{5}{4}$, and the minor third is the ratio of $\frac{3}{2}$ to $\frac{5}{4}$, namely $\frac{6}{5}$. 
Hence, starting at $C = 1$ we get $E_{\flat }= \frac{6}{5}$, $E = \frac{5}{4}$, $G = \frac{3}{2}$ and $C' = 2$. 
Similar arguments show that $F = \frac{4}{3}$, $A_{\flat} = \frac{8}{5}$ and $A= \frac{5}{3}$. 

Thus, we see that the harmonious intervals are products of powers of the prime numbers $2$, $3$ and $5$. 
The minor third is $\frac{6}{5}$, the major third is $\frac{5}{4}$, the fourth is $\frac{4}{3}$, the fifth is $\frac{3}{2}$, the minor sixth is $\frac{8}{5}$, the major sixth is $\frac{5}{3}$, and the octave is $2$. The logical deduction of such ratios is the basis of the just system of intonation, which Barbour (1953) defines as ``a system of tuning based on the octave $(2:1)$, the pure fifth $(3:2)$, and the pure major third $(5:4)$."

The situation with dissonant intervals is a little trickier. 
The tones that we have derived so far come about as a consequence of regarding $C$ as the tonal center. 
But suppose we think of $G$  as a new tonal center. 
This is like taking a new string two-thirds the length of the original and starting the analysis over.
In this way we get a value for $D$, which is a fifth above $G$. In particular, since $G_4 = \frac{3}{2}$, we multiply $\frac{3}{2}$ times $\frac{3}{2}$ to obtain $D_5 =\frac{9}{4}$ and thus $D_4 = \frac{9}{8}$. 
Since $D_4$ is a whole tone above $C_4$, we are then tempted to assign the value $\frac{9}{8}$ to the interval of a whole tone. 
Likewise, since $B$ is a major third above $G$ we multiply $\frac{3}{2}$ times $\frac{5}{4}$ to obtain $B = \frac{15}{8}$. 
And since $B_{\flat}$ is a fourth above $F$, and $F$ is a fourth above $C$, we can set $B_{\flat}$ to be equal to 
$\frac{4}{3}$ times $\frac{4}{3}$ and hence $B_{\flat}=  \frac{16}{9}$.
Finally, since $F= \frac{4}{3}$ is a semitone above $E= \frac{5}{4}$, and $C' = 2$ is a semitone above $B = \frac{15}{8}$, we conclude that the interval of a semitone can be given the ratio $\frac{16}{15}$, and hence $C_{\sharp} = \frac{16}{15}$. 
Then we can take $F_{\sharp}$ to be given by a combination of a third and a whole tone above $C$. 
This gives the product of $\frac{5}{4}$ and $\frac{9}{8}$, namely $F_{\sharp} = \frac{45}{32}$. 

So, to sum up, we have assigned rational values to all the notes of the chromatic scale given by products of powers of the prime numbers 2, 3, and 5. 
The resulting system of tuning is a good example of what we call ``just intonation" and it has many attractive features. 

In particular, the diatonic scale takes the form \eqref{just diatonic scale}. This scale has been
discussed by Gioseffo Zarlino (1517-1590) and others from the time of the Renaissance onward, though its roots lay in the work of Didymus, Ptolemy,  Porphyry and other scholars of the early Roman Empire (Barker 1984, 1994), and it is now commonly referred to as Ptolemy's intense diatonic.  For the associated chromatic scale in its entirety, according to the logical development presented thus far, we obtain \eqref{flat scale}.
Barbour (1953) refers to this version of the chromatic scale as Alexander Malcolm's Monochord, and points out that it is equivalent to Johannes Kepler's Monochord No.~2 when the latter is transposed down by a perfect fifth. We can use the term ``just intonation" to refer broadly to all of the various tuning systems constructed out of intervals involving products of powers of the primes $2$, $3$ and $5$. 
The just diatonic scale \eqref{just diatonic scale} is fairly rigid as a mathematical object in the sense that there is relatively limited scope in modifying its entries. The just chromatic scale \eqref{flat scale} is more flexible and there are many other choices that can be made for its entries, as we shall see in Section IV.

A good account of the virtues of the just system can be found in the works of  H.~C.~Longuet-Higgins (1987). It would appear that in practice the just system, whether it be in the narrow sense of the Ptolemaic and Keplerian scales mentioned above, or more broadly in the sense of the totality of all rational 5-limit dodecatonic scale systems, has served primarily as an {\it id\'ee fixe}, in relation to which a vast range of ingenious temperaments have been constructed over the last few centuries, rather than as a viable solution to the problem of keyboard tuning in itself. An exception to this principle may be that of Leonhard Euler, who (if he is to be judged on the basis of what he has said in his writings) was a staunch believer in just intonation, even if he seems to wobble on the question of whether the prime factors should be restricted to 2, 3, and 5, or extended to include 7 as well and perhaps higher primes. 
Be that as it may, there are many subtle albeit well-known issues that arise in connection with the 5-limit just system, and it is in the numerous attempts that have been made at resolving such issues that new ideas have been originated both in musical analysis and in practice. 

To get a sense of some of these issues, let us return to the idea of shifting the tonal center. 
Multiplication by $\frac{3}{2}$  leads us successively from $C_4$ to $G_4$ to $D_5$ to $A_5$ and then on to $E_6$. 
We obtain a value of $\frac{27}{8}$ for $A_5$ and hence $ \frac{81}{16}$ for $E_6$.
But that implies $E_4 = \frac{81}{64}$, whereas we had already deduced $E_4 = \frac{80}{64}$, that is to say, $\frac{5}{4}$. 
The multiplicative discrepancy $\kappa = \frac{81}{80}$ between the two values of $E_4$  is called the syntonic comma (or comma of Didymus). 
Likewise, we had set $A = \frac{5}{3} \,(= \frac{80}{48})$, whereas now we get $A = \frac{27}{16}\,(= \frac{81}{48})$, again with a discrepancy of $\frac{81}{80}$. 
The just system is evidently unstable under such shifts in the tonal center. 

Going forward, let us write $P/Q$ for the ratio of the tones $P$ and $Q$. This ratio represents the interval from the tone $Q$ to the tone $P$ (often written $Q$\,--\,$P$). For example, the interval from $E = \frac{5}{4}$ to $G = \frac{3}{2}$ is the ratio
$G/E = \frac{3}{2} \div \frac{5}{4} = \frac{6}{5}$, a minor third.

 If we recall that our value for $D$ was obtained by applying the cycle of fifths to $G$, we have to ask whether $D = \frac{9}{8}$ is consistent with $A = \frac{5}{3}$.  
In fact, one finds that  the interval $A/D$ is a bit flat for a fifth  if $D = \frac{9}{8}$. 
We obtain 
\begin{eqnarray}
A/D = \tfrac{5}{3} \div \tfrac{9}{8} = \tfrac{40}{27} = \kappa^{-1} \, \tfrac{3}{2}.
\end{eqnarray} 
 Since $\kappa > 1$, we see that the interval $A/D$ obtained in this way is slightly smaller than $\frac{3}{2}$, i.e.~somewhat flat. To sensitive ears it sounds out of tune, particularly if played on the harpsichord, and hence has come to be known as an example of a ``wolf\," interval. 
But if we set $D = \frac{10}{9}$,  the resulting interval 
\begin{eqnarray}
A/D = \tfrac{5}{3} \div \tfrac{10}{9} = \tfrac{3}{2}
\end{eqnarray} 
is a perfect fifth. 
Indeed, one can check that $\frac{9}{8} \div \frac{10}{9} = \kappa$. 
Thus, it seems inevitable that we must augment the just system with a second version of the note $D$ taking the value $\frac{10}{9}$. 

Similarly, the value $B_{\flat} = \frac{16}{9}$ that we tentatively assigned earlier  is consistent with the idea that 
$B_{\flat}$ should be a fourth above $F$. 
On the other hand, $B_{\flat }$ can also be regarded as a minor third above $G$, or a fifth above $E_{\flat}$ and these both give $B_{\flat} = \frac{9}{5}$. 
Thus, we need to assign two different values to $B_{\flat }$, namely $\frac{16}{9}$ and $\frac{9}{5}$, and we can check that $\frac{9}{5} \div \frac{16}{9} = \kappa$. Indeed, the situation of $B_{\flat }$ precisely parallels that of $D$.

Likewise, one can show that four different values for $F_{\sharp }$ are required, corresponding to the various ways that the interval $F_{\sharp }/C$ can be formed out of simpler intervals. 
These are 
\begin{eqnarray} \label{F sharp}
F_ {\sharp} = \tfrac{45}{32}, \quad F_ {\sharp} = \tfrac{64}{45}, \quad F_ {\sharp} = \tfrac{25}{18}, \quad F_ {\sharp} = \tfrac{36}{25}. 
\end{eqnarray} 
Note that
$\frac{45}{32} \div \frac{25}{18} = \kappa$ and that  $\frac{45}{32}$ and $\frac{64}{45}$ are complements in the sense that their product is the octave.   The multiplicity of the intervals representing $F_ {\sharp}$ is troubling, and all are rather complicated when they are factorized into products of powers of primes.
We call this ``the problem of $F_{\sharp}"$. Is there a way around the problem of 
$F_{\sharp}$\,? This question motivates us both to revisit the foundations of the just system and to introduce the idea of complex tonality.

%
\section{Atoms, Molecules and Particles}
\noindent To enable a way forward, leading to the complex systems that we introduce later in the article, it will be helpful first if we look at the Ptolemaic system in terms of so-called ``atoms".
The idea is that we introduce a collection of simple tones (called atoms) such that each tone in the diatonic scale can be regarded as a multiplicative aggregation of atoms into molecules. 
The notes of the just system can be built up in this way step by step. 
We introduce three atoms, given by 
\begin{eqnarray}
X = \tfrac{9}{8}, \, \, Y = \tfrac{10}{9}, \, \,Z = \tfrac{16}{15}. 
\end{eqnarray} 
Then one sees that 
\begin{eqnarray}
XY = \tfrac{5}{4}, \, \, XYZ = \tfrac{4}{3}, \, \, X^2YZ = \tfrac{3}{2}, \, \, X^2Y^2Z = \tfrac{5}{3}, \, \, X^3Y^2Z = \tfrac{15}{8}, \, \, X^3Y^2Z^2 = 2, 
\end{eqnarray} 
and hence that for the $C$-major scale in the just system we can write
\begin{eqnarray}
&&
\quad \quad \quad \, \, C = 1, \, \,
D = X, \, \,
E = XY, \, \,
F = XYZ, \, \,
\nonumber \\ && G = X^2YZ, \, \,
A = X^2Y^2Z, \, \,
B = X^3Y^2Z, \, \,
C' = X^3Y^2Z^2. 
\end{eqnarray} 
Each note in the scale is represented by a ``molecule" with a particular configuration of atoms, and as the scale progresses more atoms are aggregated. Our atoms consist of the two ``whole tones", viz., $\tfrac{9}{8}$ and $\tfrac{10}{9}$, alongside the semitone $\tfrac{16}{15}$. We can use a shorthand to indicate the order in which the atoms are added in.
Thus, for the diatonic scale starting at $C = 1$ we have the {\it aggregation sequence} $1: X, Y, Z, X, Y, X, Z$. The note before the colon denotes the initial tone, which is followed by a sequence of atoms.
The initial unity here ensures that the scale begins at our chosen value for $C$, but in principle one can choose a different initial value. The algebraic identity $X^3Y^2Z^2 = 2$ that closes the octave implies some flexibility in the order in which the atoms can be aggregated. The seven ecclesiastical modes arise in this way, for example, corresponding to permutations of the original aggregation sequence:    
\vspace{0.4 cm}
\\ \vspace{0.3 cm}
\quad Ionian mode: $C, D, E, F, G, A, B, C'$ $\sim 1 : X, Y, Z, X, Y, X, Z$
\\ \vspace{0.3 cm}
\quad Dorian mode: $C, D, E_\flat, F, G, A, B_\flat, C'$ $\sim 1 : X, Z, Y, X, Y, Z, X$
\\ \vspace{0.3 cm}
\quad Phrygian mode: $C, D_\flat, E_\flat, F, G, A_\flat, B_\flat, C'$ $\sim 1 : Z, X, Y, X, Z, Y, X$
\\ \vspace{0.3 cm}
\quad Lydian mode: $C, D, E, F_\sharp, G, A, B, C'$ $\sim 1 : X, Y, X, Z, Y, X, Z$
\\ \vspace{0.3 cm}
\quad Mixolydian mode: $C, D, E, F, G, A, B_\flat, C'$ $\sim 1 : X, Y, Z, X, Y, Z, X$
\\ \vspace{0.3 cm}
\quad Aeolian mode: $C, D, E_\flat, F, G, A_\flat, B_\flat, C'$ $\sim 1 : X, Z, Y, X, Z, Y, X$
\\ \vspace{0.3 cm}
\quad Locrian mode: $C, D_\flat, E_\flat, F, G_\flat, A_\flat, B_\flat, C'$ $\sim 1 : Z, X, Y, Z, X, Y, X$.

One can think of the note representing the underlying key (in this case, $C$) as being given in the form of a structureless monad, represented by the tone unity. But set in a larger context, this note itself can be given structure.  It may be helpful to introduce a chemical analogy. The note $C$ corresponds to the completed shell corresponding to an inert gas, such as argon. The creation of a scale, in this analogy, is represented by the addition of electrons, one by one, until the next shell is complete, corresponding to the note $C'$. 

It is interesting to note that Isaac Newton seems to have been in favour of the just system but preferred a version of the scheme based on the Dorian mode on account of the symmetric arrangement of the atoms arising in that case (Bibby 2003). 
It may be that Newton was attracted to the fact that the aggregation sequence forms a palindrome in this mode. In any case it would be interesting to pursue the theory of modes arising in the present context from the point of view of the algebraic combinatorics of words, in the spirit of Dominguez, Clampitt and Noll  (2009), $\check{\rm{Z}}$abka (2011), Clampitt (2017), Clampitt and Noll (2018), Clampitt (2019), and others.  

Inspection of the modal scales leads to another method for arriving at the accidentals of the chromatic scale. The resulting values for $D_\flat $, $E_\flat $, $A_\flat$, and $B_\flat$ are consistent with the values we had tentatively recorded at \eqref{flat scale}.  We obtain $B_\flat= XZYXYZ = \frac{16}{9}$ for the Dorian mode, $B_\flat= ZXYXZY = \frac{16}{9}$ for the Phrygian, 
$B_\flat= XYZXYZ = \frac{16}{9}$ for the Mixolydian, $B_\flat= XZYXZY = \frac{16}{9}$ for the Aeolian, and $B_\flat= ZXYZXY = \frac{16}{9}$ for the Locrian.

Likewise, we obtain $E_\flat = XZ = \frac{6}{5}$ for the Dorian and Aeolian modes, $E_\flat = ZX = \frac{6}{5}$ for the Phrygian and Locrian modes, $A_\flat= ZXYXZ = \frac{8}{5} $ for the Phrygian,  
$A_\flat= XZYXZ = \frac{8}{5}$ for the Aeolian, $A_\flat= ZXYZX = \frac{8}{5}$ for the Locrian,  and $D_\flat = Z = \frac{16}{15}$ for the Locrian. But we get $F_\sharp = XYX = \frac{45}{32}$ in the Lydian mode and $G_\flat = ZXYZ = \frac{64}{45}$ in the Locrian mode. So the problem of $F_\sharp$ rears its head, since the modes give two different values for it.  

The modal scales as we have presented them are all rooted in the same tonality, that of $C$. 
If we change the tonal center, instabilities emerge in the form of new tones. 
Suppose, for example, we wish to make a selection of notes from the just chromatic scale to represent the notes of the $G$-major scale. 
As with the $C$-major diatonic scale, we can ask that the $G$-major scale should be constructed via the accumulation of atoms. This can be achieved as follows. 
We begin with the molecule for $G_4$, which we know to be of the form $X^2YZ$. 
Then we add on successively the atoms $Y, X, Z, X, Y, X, Z$. The resulting tones correspond to existing notes on the just keyboard.

At the next step, problems begin to emerge.
If we try to construct a $D$-major scale, then we can start at $D = \frac{9}{8}$ and add successive atoms by applying the sequence
$Y, X, Z, Y, X, X, Z$. 
We obtain $F_{\sharp} = \frac{45}{32}$, which we have agreed is acceptable, but for $C_{\sharp 5}$ we obtain $\frac{135}{64}$, which is not one of the pre-existing tones. 

When we go on to look at the remaining sharp keys we obtain three more tones, given by $G_{\sharp} = \frac{25}{16}$, $D_{\sharp} = \frac{75}{32}$, and $A_{\sharp}= \frac{225}{128}$, contradicting the values previously assigned. 
The flat keys fare better, but in the key of $D_\flat$ major we obtain  $G_{\flat} = \frac{64}{45}$, which is different from the $F_{\sharp } = \frac{45}{32}$ we agreed on earlier. 
And in the key of $G_{\flat}$ major, we obtain $C_{\flat }= \frac{48}{25}$, which is different from the previously proposed tuning at $B = \frac{15}{8}$. 
Apparently, if we try to represent the major scales by successive aggregation of the atoms proposed to construct the just diatonic scale, then we obtain new values for notes to which we have already tentatively assigned values.

The generally accepted way around these difficulties in practice is, of course, the system of {\em equal temperament}, in which we introduce a irrational semitone 
\begin{eqnarray}\psi = \sqrt[12]{2},
\end{eqnarray}
which functions as the single ``particle" out of which all the tones of the chromatic scale can be assembled by aggregation. 
Thus for the equally tempered chromatic scale we obtain
\begin{eqnarray}
&&
C = 1, \, \,
C_{\sharp} =  \psi,  \, \,
D = \psi^2, \, \,
D_ {\sharp}=  \psi^3, \, \,
E = \psi^4,  \, \,
F = \psi^5,  \, \,
F_ {\sharp} = \psi^6,  \, \,
\nonumber \\ && \quad  G = \psi^7,  \, \,
 G_ {\sharp} = \psi^8,  \, \,
A = \psi^9,  \, \,
A_ {\sharp} = \psi^{10},  \, \,
B = \psi^{11},  \, \,
C' =  \psi^{12}.
\end{eqnarray} 
and we have the aggregation sequence 
\begin{eqnarray}
1: \psi,  \psi^2, \psi^3,  \psi^4, \psi^5, \psi^6, \psi^7,  \psi^8, \psi^9,  \psi^{10}, \psi^{11},  \psi^{12}. 
\end{eqnarray} 
Note that $F_ {\sharp} = \sqrt{2}$, and that $C' = 2$. Under equal temperament none of the harmonious intervals are perfect. 
Nevertheless, overall, the system produces satisfactory results for music written to be played in that system. 
This includes, for example, a good deal of the repertoire of music written for the piano.
Equal temperament thus offers a tentative resolution to the paradox of tuning by spreading the dissonance equally across the keyboard. Scholes (1955) goes further in his {\em Oxford Companion to Music} at page 1018, and says: 
\begin{quote} \begin{small}The whole development of musical composition in the nineteenth century was based on the acceptance of the twelve equal keys into which the composer could pass without let or hindrance. Wagner's chromatic harmonies, for example, necessarily imply an orchestra with its instruments tuned `equally'. \end{small}
\end{quote}
Opponents of equal temperament, of which there are many,  claim that it flattens out the music, leaving it homogeneous and characterless. Worse still, although the fifths are reasonably accurate under equal temperament, the thirds are noticeably wide.
The disadvantages have been succinctly summarized by Klopp (1974): 
\begin{quote} \begin{small} Obviously, all keys are available and the possibilities of enharmonization are unlimited. However, these freedoms are dearly paid for: there is not one single pure interval; the thirds are especially poor, giving the triad an insecure and restless sound; there is no differentiation between the keys; melodic tensions are reduced; the temperament is difficult to set. \end{small}
\end{quote}
And according to Tovey (1929):
\begin{quote} \begin{small} No true harmonic ideas are based on equal temperament.\end{small}
\end{quote}

Euler (1739, 1774) was dead against equal temperament for essentially the same reasons and expressed the opinion that the majority of musicians had rejected it. But in this he seems to have been mistaken, and it is clear that he did not anticipate the great wave of nineteenth and twentieth century composers who would embrace the advantages that transpositions into any key would offer. Nonetheless, one can take the view that from the Age of Enlightenment onward the ever-increasing demand for modulations into keys far removed from the tonic pushed the principles of Western tonal music as far as they could be taken---and that despite the fact that music predicated on equal temperament has produced numerous striking, important, and completely convincing compositions, there remains a lingering sense of incompleteness in the centuries-old mathematical project of developing a system of music based entirely on products of powers of primes, in the spirit of Euler's explorations. 
This suggests that we should return to the idea of a tonal system based on the rationals and consider in more precise terms the conditions that we would like it to satisfy, with an eye on the possibility of novel generalizations.

%
\section{Chromatic tone systems}

\noindent  Let us enquire, therefore, whether it is possible to create a version of the just chromatic scale by the successive aggregation of ``particles" (semitones of various magnitudes) in such a way that the result includes all of the notes of the just diatonic scale as a subscale. Although not, perhaps, obvious, the answer is affirmative, as the following example (Orbach 1999), based on Kepler's scale, clearly illustrates. As particles we take
\begin{eqnarray}\label{historic scheme}
p = \tfrac{16}{15}, \quad q = \tfrac{135}{128}, \quad r = \tfrac{25}{24}. 
\end{eqnarray} 
An aggregation sequence that leads to the just chromatic scale \eqref{flat scale} is then given as follows: 
\begin{eqnarray} \label{aggregation order}
1: p, \,q, \, p, \, r, \, p, \, q, \, p, \, p, \, r, \, p, \, q, \, p.
\end{eqnarray} 
That this sequence reproduces the atomic sequence $1: X, Y, Z, X, Y, X, Z$ for the just diatonic scale with $X = \tfrac{9}{8}$, $Y = \tfrac{10}{9}$ and $Z = \tfrac{16}{15}$ can easily be verified if one observes that $X = pq$, $Y = pr$ and $Z = p$. But is this the only chromatic scale that can be constructed from three rational constituents by successive aggregation in such a way that it includes the just diatonic scale as a subscale? 

The answer, perhaps surprisingly, turns out to be negative: in fact, there are ninety-six different versions of the chromatic scale that interleave the just diatonic scale. The situation is summarized  in the following.

\begin{Theorem}
Let $S$ denote a set of three rational semitones with the property that there exists a chromatic scale of twelve notes constructed by aggregation from elements of $S$ interleaving the just diatonic scale $C =1$, $D = \frac{9}{8}$, $E = \frac{5}{4}$, $F = \frac{4}{3}$,  $G =  \frac{3}{2}$, $A = \frac{5}{3}$,  $B= \frac{15}{8}$, $C' = 2$.  Three and only three such sets of semitones exist, given by

\vspace{.25cm}
{\rm Just System I:} $S = \{ \tfrac{16}{15}, \hspace{.02cm} \tfrac{135}{128}, \hspace{.02cm} \tfrac{25}{24}\}$, 

\vspace{.25cm}
{\rm Just System  II:} $S = \{ \tfrac{135}{128}, \hspace{.02cm} \tfrac{16}{15}, \hspace{.02cm}  \tfrac{256}{243}\}$,

\vspace{.25cm} 
{\rm Just System III:}   $S = \{  \tfrac{25}{24}, \hspace{.02cm} \tfrac{27}{25}, \hspace{.02cm}\tfrac{16}{15} \}$. 

\vspace{.25cm}
\noindent Each such system  gives rise to thirty-two distinct chromatic scales.
\end{Theorem}
\noindent \textit{Proof}. To construct a twelve-tone chromatic scale that interleaves the just diatonic scale, we need to assemble the atoms $X = \tfrac{9}{8}$, $Y = \tfrac{10}{9}$, $Z = \tfrac{16}{15}$ from a set of three particles $\{p, q, r \}$ in such a way that $X$ and $Y$ each contain two particles, and $Z$ contains a single particle. Clearly, $X$ and $Y$ share a particle in common, say, $p$. Then without loss of generality we can set $X = pq$ and $Y=pr$, and $Z$ will take one of the values $p$, $q$, $r$. 
Thus, $pq = \tfrac{9}{8}$ and $pr = \tfrac{10}{9}$, and either $p = \tfrac{16}{15}$ (System 1), or $q = \tfrac{16}{15}$ (System 2), or  $r = \tfrac{16}{15}$ (System 3). Solving for $p$, $q$, $r$, we obtain the three systems stated in the theorem. In any one of the systems, the order of aggregation for the constituents of $X$ and the constituents of $Y$ can be interchanged for each $X$ and each $Y$. Since there are three $X$s and two $Y$s, that gives us five possible interchanges, and hence thirty-two different variations for each system. 
\qed 
\vspace{0.5cm}

Just System I is the historic scheme of semitones set out in \eqref{historic scheme}. Within any of the three systems, we can present the constituents of $X$ either in the order $p, q$ or the order $q, p$. Likewise the constituents of $Y$ can be presented in the order $p, r$ or the order $r, p$.  Thus for each of the five ``flat" notes of the chromatic scale, we have two (and only two) choices. 

For example, in System I we obtain either $\tfrac{16}{15}$ or $\tfrac{135}{128}$ for $D_\flat$,  either $\tfrac{6}{5}\,(= \tfrac{9}{8}\cdot \tfrac{16}{15})$ or $\tfrac{75}{64} \, (= \tfrac{9}{8}\cdot \tfrac{25}{24})$ for $E_\flat$, either 
$\tfrac{45}{32} \, (= \tfrac{4}{3}\cdot \tfrac{135}{128})$ or $\tfrac{64}{45} (= \tfrac{4}{3}\cdot \tfrac{16}{15})$ for $G_\flat$, either $\tfrac{8}{5} \, (= \tfrac{3}{2}\cdot \tfrac{16}{15})$ or $\tfrac{25}{16} \, (= \tfrac{3}{2}\cdot \tfrac{25}{24})$ for $A_\flat$, and either $\tfrac{16}{9}\, (= \tfrac{5}{3}\cdot \tfrac{16}{15})$ or 
$\tfrac{225}{128}\, (= \tfrac{5}{3}\cdot \tfrac{135}{128})$ for $B_\flat$. Thus we have thirty-two different versions of System I, depending on which choices we make for the five accidentals. 

If we take the first option in each accidental, for which the tones are the simplest, then we recover the  aggregation sequence \eqref{aggregation order}, which under System I gives rise to \eqref{flat scale}, which appears in Barbour (1953) as the historic tuning system known as Malcolm's Monochord. Barbour gives the tones in cents, so one has to translate these approximate cent values to the corresponding exact rational numbers to obtain our chromatic scale \eqref{flat scale}. By \eqref{historic scheme} and \eqref{aggregation order}, the aggregation sequence for Malcolm's Monochord is given more explicitly by 
%
%
%
\begin{eqnarray}
1: \tfrac{16}{15},\,  \tfrac{135}{128},\, \tfrac{16}{15},\, \tfrac{25}{24},\, \tfrac{16}{15},\, \tfrac{135}{128},\, \tfrac{16}{15},\, \tfrac{16}{15},\, \tfrac{25}{24},\,  \tfrac{16}{15},\, \tfrac{135}{128},\, \tfrac{16}{15}, 
\end{eqnarray} 

\noindent starting at $C$. Now let's consider Kepler's Monochord No.~2, which is given by the scale 
%
%
\begin{eqnarray}
&&
C = 1, \, \,
D_{\flat} = \tfrac{135}{128}, \, \,
D = \tfrac{9}{8}, \, \,
E_ {\flat}= \tfrac{6}{5}, \, \,
E = \tfrac{5}{4},  \, \,
F = \tfrac{4}{3},  \, \,
G_ {\flat} = \tfrac{45}{32},  \, \,
\nonumber \\ && \quad \, \, G = \tfrac{3}{2},  \, \,
 A_ {\flat} = \tfrac{8}{5},  \, \,
A = \tfrac{27}{16},  \, \,
B_ {\flat} = \tfrac{9}{5},  \, \,
B = \tfrac{15}{8},  \, \,
C' = 2.
\end{eqnarray} 
This scale as such does not belong to Just System I, since $A = \tfrac{27}{16}$ rather than $\tfrac{5}{3}$. But if we transpose Kepler's tuning down by a fifth, along the whole keyboard, thus assigning $F_3 = \tfrac{2}{3}$, $F_{\sharp 3} = \tfrac{2}{3}\cdot \tfrac{135}{128} = \tfrac{45}{64}$, $G_3 = \tfrac{2}{3}\cdot \tfrac{9}{8} = \tfrac{3}{4}$, and so forth, then we recover the chromatic scale \eqref{flat scale}, as we mentioned earlier. Hence Kepler's system is isomorphic to that of Malcolm, and admits the same aggregation sequence of semitones once a suitable starting point has been established following a transposition. 

It may be helpful if we set out the structure of the three systems in a little more detail, showing the breakdown of composite atoms $X$ and $Y$ into their particle constituents:

\vspace{.25cm}
Just System I: \, $X =  \tfrac{16}{15} \cdot  \tfrac{135}{128}$, \,  $Y = \tfrac{16}{15} \cdot  \tfrac{25}{24}$,

\vspace{.25cm}
Just System II: \, $X =  \tfrac{135}{128} \cdot  \tfrac{16}{15}$,  \, $Y = \tfrac{135}{128} \cdot  \tfrac{256}{243}$,

\vspace{.25cm}
Just System III: \, $X =  \tfrac{25}{24} \cdot  \tfrac{27}{25}$,  \, $Y = \tfrac{25}{24} \cdot  \tfrac{16}{15}$.

\vspace{.25cm}
\noindent Note that the intervals arising here are well-established in the theory of tone systems, under a variety of names; see, e.g., Helmholtz (1885), Longuet-Higgens (1987), Halu$\check{\rm{s}}$ka (2004). We have met the major whole tone $\tfrac{9}{8}$, the minor whole tone 
$\tfrac{10}{9}$, and the (minor) diatonic semitone $\tfrac{16}{15}$. Now we meet the  minor chromatic semitone $\tfrac{25}{24}$, 
the major chromatic semitone $\tfrac{135}{128}$, the major diatonic semitone $\tfrac{27}{25}$, and the minor Pythagorean semitone $\tfrac{256}{243}$.

We observe that various relations involving multiplication by the syntonic comma $\kappa = \frac{81}{80}$ hold between the semitone intervals determined by Theorem 1. In particular, we have
\begin{eqnarray}
\tfrac{256}{243} \xlongrightarrow{\kappa} \tfrac{16}{15} \xlongrightarrow{\kappa} \tfrac{27}{25}, 
\quad \quad \tfrac{25}{24} \xlongrightarrow{\kappa} \tfrac{135}{128}.
\end{eqnarray} 
What this means is that a set of discrete transformations can be constructed taking the three systems from one to another. Let us write 
\begin{eqnarray}\label{five semitones}
\theta_{-1} = \tfrac{256}{243},\quad \theta_{0} = \tfrac{16}{15},  \quad \theta_{1} = \tfrac{27}{25}, \quad
\phi_0 = \tfrac{25}{24}, \quad \phi_1 = \tfrac{135}{128}.
\end{eqnarray} 
Then one can regard the three $\theta$ semitones as being different versions of a common underlying entity, differing by factors of the syntonic comma, and similarly for the two $\phi$ semitones. One can think of 
$\theta_{0}$ as being a kind of ``neutral" semitone, whereas $\theta_{1}$ has a ``charge" of $1$ and $\theta_{-1}$ has a charge of $-1$. Similarly, $\phi_0$ is neutral and  $\phi_1$ has charge one. 

Then the relations between the different systems amount to transferring units of charge from one semitone to another in such a away that the total charges of $X$ and $Y$ remain unchanged. Thus $X$ is a charged whole tone, $Y$ is a neutral whole tone, $Z$ is a neutral semitone, and the three just tuning systems can be set out in the following scheme: 

\vspace{.25cm}
Just System I: \, $X =  \theta_{0} \cdot \phi_1$, \,  $Y = \theta_{0} \cdot  \phi_0$, \, $Z = \theta_{0}$,

\vspace{.25cm}
Just System II: \, $X =  \theta_{0} \cdot \phi_1$, \,  $Y = \theta_{-1} \cdot  \phi_1$, \, $Z = \theta_{0}$,

\vspace{.25cm}
Just System III: \, $X =  \theta_{1} \cdot \phi_0$, \,  $Y = \theta_{0} \cdot  \phi_0$, \, $Z = \theta_{0}$.

\vspace{.25cm}
\noindent Thus in moving from System I to System II, we transfer a unit of charge from $\theta_0$ to $\phi_0$ in the $Y$ atom; moving from System I to System III we transfer a unit of charge from $\phi_1$ to $\theta_0$ in the $X$ atom;  whereas in moving from System II to System III we transfer a unit of charge from $\phi_1$ to $\theta_0$ in the $X$ atom and we transfer a unit from $\phi_1$ to $\theta_{-1}$ in the $Y$ atom. We see that the three systems are closely related and that there is order to the arrangement. 

\vspace{.25cm}

\noindent {\bf Remark 1}. The existence of a five-semitone chromatic scale agreeing with Ptolemy's intense diatonic scale on the notes of that scale can be demonstrated by use of the various semitones appearing
 in \eqref{five semitones}. This can be achieved by setting
\begin{eqnarray}
X =  \theta_{1} \cdot \phi_0, \quad Y = \theta_{-1} \cdot  \phi_1, \quad Z = \theta_{0}
\end{eqnarray} 
\noindent and using the chromatic aggregation sequence
\begin{eqnarray}
1:  \theta_{1},  \,\phi_0, \,\theta_{-1},\, \phi_1,\, \theta_{0}, \,\theta_{1}, \, \phi_0,\, \theta_{-1}, \,\phi_1,\, \theta_{1}, \, \phi_0,  \,\theta_{0}.
\end{eqnarray} 

An example of a historical scale that (modulo transposition) belongs to Just System II can be found in the work of Bartolomeus Ramis de Pareja (1482). Ramis's Monochord is given by Barbour (1953) both in cents and in the Eitz notation, i.e.~in Pythagorean tunings that may be adjusted up or down by one or more syntonic commas.  Expressed in rationals, Ramis's tuning starting at $C$ takes the form
%
%
%
\begin{eqnarray}
&&
C = 1, \, \,
D_{\flat} = \tfrac{135}{128}, \, \,
D = \tfrac{10}{9}, \, \,
E_ {\flat}= \tfrac{32}{27}, \, \,
E = \tfrac{5}{4},  \, \,
F = \tfrac{4}{3},  \, \,
G_ {\flat} = \tfrac{45}{32},  \, \,
\nonumber \\ && \quad \, \, G = \tfrac{3}{2},  \, \,
 A_ {\flat} = \tfrac{128}{81},  \, \,
A = \tfrac{5}{3},  \, \,
B_ {\flat} = \tfrac{16}{9},  \, \,
B = \tfrac{15}{8},  \, \,
C' = 2.
\end{eqnarray} 

\noindent This is clearly not within the scope of Theorem 1, since $D = \tfrac{10}{9}$. Nonetheless, the associated aggregation sequence, which is of the form
%
%
\begin{eqnarray}
1: \tfrac{135}{128},\, \tfrac{256}{243},\,
\tfrac{16}{15},\, \tfrac{135}{128},\, \tfrac{16}{15},\, \tfrac{135}{128}, \,\tfrac{16}{15}, \,\tfrac{256}{243},\,
\tfrac{135}{128}, \, \tfrac{16}{15},\, \tfrac{135}{128},\, \tfrac{16}{15},
\end{eqnarray} 
\noindent exhibits the particle structure of System II, which suggests, as we saw in the case of Kepler's scale, that a transposition might bring it into line with Theorem I. Indeed, this turns out to be the case, for if we transpose the scale of Ramis de Pareja down by a fourth, multiplying each note by $\tfrac{3}{4}$ all along the keyboard, then starting at $C$, we obtain 
%
\begin{eqnarray}
&&
C = 1, \, \,
D_{\flat} = \tfrac{135}{128}, \, \,
D = \tfrac{9}{8}, \, \,
E_ {\flat}= \tfrac{32}{27}, \, \,
E = \tfrac{5}{4},  \, \,
F = \tfrac{4}{3},  \, \,
G_ {\flat} = \tfrac{45}{32},  \, \,
\nonumber \\ && \quad \, \, G = \tfrac{3}{2},  \, \,
 A_ {\flat} = \tfrac{405}{256},  \, \,
A = \tfrac{5}{3},  \, \,
B_ {\flat} = \tfrac{16}{9},  \, \,
B = \tfrac{15}{8},  \, \,
C' = 2,
\end{eqnarray} 
\noindent  which evidently satisfies the conditions of Theorem 1; and for the associated aggregation sequence starting at $C=1$, which is consistent with Just System II, we have
\begin{eqnarray}
1: \tfrac{135}{128},\,
\tfrac{16}{15},\, \tfrac{256}{243},\, \tfrac{135}{128},\, \tfrac{16}{15},\, \tfrac{135}{128}, \,\tfrac{16}{15}, \,
\tfrac{135}{128}, \,\tfrac{256}{243},\, \tfrac{16}{15},\, \tfrac{135}{128},\, \tfrac{16}{15}.
\end{eqnarray} 
Historical examples of scales belonging to Just System III that can be found (without need of transposition) in Barbour's list include Fogliano's Monochord No.~2, and Rousseau's Monochord (Fogliano 1529, Rousseau 1768). But there are other historical just tunings in Barbour's list that are of interest, some involving as many as four semitones, such as Euler's Monochord, that would also be worthy of study using the methods we have proposed. 

Summing up, we have seen that in the just system the notes of various well-established scales can be assembled from a small number of semitone particles, and we have been able to present a more or less complete account of the tuning schemes that can be constructed on the premise that the structure of the Ptolemaic scale is preserved and that the number of semitones is three. 

Nevertheless, we are left with the uncomfortable fact that the resulting systems of semitones involve products of rather high powers of prime numbers. Is it possible to improve on the situation in some way, by introducing a new scheme for music that allows one to bring the multiplicities of the primes down, so we are only left with {\it low} powers? This is the question we seek to answer in our investigation of complex intonation.
%
%
\section{Complex diatonic scales}\label{section V}
\noindent We shall give an interesting example of such a scheme. 
It goes outside of the usual conventions of music by allowing tones to be represented by {\em complex numbers}. 
We begin with the observation that all the intervals under consideration in the just system are products of powers of the numbers 2, 3 and 5. 
These numbers are primes in the conventional sense, but 2 and 5 can be factorized over the complex numbers. That is to say, we have
\begin{eqnarray}
2 = (1+ \ri) (1 - \ri) \quad \text{and} \quad
5 = (1+ 2\ri) (1 - 2\ri). 
\end{eqnarray}
In particular, since the note $F_\sharp$ lies midway between $C$ and $C'$, this suggests we should set 
\begin{eqnarray}
F_\sharp = 1+ \ri. 
\end{eqnarray} 

It may  not be obvious \hspace{-.2cm} {\em a priori} what kind of acoustic event will trigger the sensation of a complex tone, but since the numbers involved are simple and striking, we are encouraged to explore further, adding further complex tones. 

First, we shall pursue the matter informally, as we did in our initial analysis of the just system, following which we presented Theorem 1, which gave a precise account of the system in its entirety. Thus here we shall construct an example of a complex scheme by use of clever albeit essentially inductive reasoning. Then once we have by this line of argument shown the existence of a viable complex scheme we proceed to formalize the theory via the presentation of Theorem 2. 
 
Let us write $\xi$ for $1+ \ri$. Since an octave is composed of two tritones, this suggests that we should identify both
$1+ \ri$ and $1- \ri$ as representations of the tritone. As we remark later, our suggestion that $1+ \ri$ and $1- \ri$ should be regarded as representing the same tone can be understood as a reflection of the fact that they differ by a ``unit", in this case an overall factor of $\ri$. As basic intervals we retain the diatonic whole tones $X = \frac{9}{8}$ and $Y = \frac{10}{9}$, and the semitone $Z = \frac{16}{15}$, and supplement these with a complex semitone $\alpha$ defined by 
\begin{eqnarray}
\alpha = \tfrac{3}{4}\, \xi
\end{eqnarray} 
and a complex whole tone $\delta$ defined by 
\begin{eqnarray}
\delta = \tfrac{4}{5}\,\xi.
\end{eqnarray} 
The argument is that since $\frac{4}{3} \,\alpha = \xi$ we can say that $\xi$ is a fourth above $\alpha$. Similarly, we have $\frac{5}{4}\,\delta = \xi$, so we can say that $\xi$ is a third above $\delta$.
Note also that  $\xi \,\bar\xi = 2$, $\alpha \bar \alpha = X$, $\delta = \alpha Z$  and $\alpha \bar \delta = XZ$. These various relations justify our interpretation of $\alpha$ as a type of semitone and $\delta$ as a type of whole tone. 

In our complex scheme we keep the just values for $E = \frac{5}{4}$, $F = \frac{4}{3}$, $G=\frac{3}{2}$, and $A=\frac{5}{3}$. Then in addition to $F_\sharp = \xi$ we can set $C_\sharp = \frac{3}{4}\xi$, $D_\sharp = \frac{5}{6}\xi$, and $A_\sharp = \frac{5}{4}\xi$, 
since these notes are  respectively a fourth below $F_\sharp$, a minor third below $F_\sharp$, and a third above $F_\sharp$.
We also need values for $D$, $G_{\sharp}$ and $B$. 
If we set $D = \frac{9}{8}$ as one does in just tuning, then although it is a fourth below $G$, it is a ``wolf\," fifth below $A$, since $A/D =  \frac{5}{3} \div  \frac{9}{8} = \frac{40}{27}$. But if we set $D = \frac{10}{9}$, then $D$ would lie a  fifth below $A$, but would be a ``wolf\," fourth below $G$, since
 $\frac{3}{2} \div  \frac{10}{9} = \frac{27}{20}$.
Clearly, neither of the choices $D = \frac{9}{8}$ and $D = \frac{10}{9}$ are satisfactory. 
 
 The resolution of this dilemma brings in the other complex factorization that we have at our disposal, namely 
  $5 = (1+ 2\ri) (1 - 2\ri)$. 
 Then, in addition to the complex semitone $\alpha = \frac{3}{4}\xi$, which has the property that $\alpha \bar\alpha = \frac{9}{8}$, we obtain another complex semitone $\beta$, defined by 
\begin{eqnarray}
\beta = \tfrac{1}{3}(3 + \ri),
\end{eqnarray} 
with the property that  $ \beta \bar\beta =  \frac{10}{9}$. Then we can set $G_{\sharp} = \frac{1}{2}(3 - \ri)$, a semitone above $G$. 
We observe that both of the diatonic whole tones factorize,
 which justifies the interpretation of $\beta$ as another species of semitone.
 The product of these complex semitones gives a complex whole tone which we define by setting $\alpha \beta = \gamma$, and one can check that 
\begin{eqnarray}
\gamma =  \tfrac{1}{2} (1 + 2\ri).
\end{eqnarray} 
Note that $\gamma \bar \gamma =  \frac{5}{4}$, so $\gamma$ lies midway between $C$ and $E$ in the  sense that $\xi$ lies midway in the range between the notes $C$ and $C'$. 
Thus, we set $D = \gamma$, banishing the wolves by moving $D$ off into the complex. 

We observe that the intervals $G/D = \frac{3}{5} (1-2\ri)$ and $A/D = \frac{2}{3} (1-2\ri)$ are products of low powers of Gaussian primes and hence {\em both} can be regarded as consonant in our scheme.
These intervals can be contrasted with the wolves $\frac{40}{27}$ and $\frac{27}{20}$ that arise in the just scheme.

Another unsatisfactory aspect of the just scheme is the value of $B = \frac{15}{8}$. 
This gives us $B/F = \frac{45}{32}$, the same harsh interval that we eliminated by setting $F_{\sharp} = \xi$. 
That suggests that instead of $B = \frac{15}{8}$ we should set $B = \frac{4}{3} \xi$. 
Then we obtain the simple ratio $B/F = \xi$, which by virtue of the fact that it involves only a single prime can  be regarded as consonant. 

Summing up, with the various adjustments in place, we conclude that the $C$-major scale in our complex scheme takes the form
\begin{eqnarray} 
C = 1, \, \,
D = \gamma, \, \,
E = \tfrac{5}{4}, \, \,
F = \tfrac{4}{3}, \, \,
G = \tfrac{3}{2}, \, \,
A = \tfrac{5}{3}, \, \,
B = \tfrac{4}{3} \xi, \, \,
C' = 2,  
\end{eqnarray} 
with the aggregation sequence $1: \gamma, \bar\gamma, Z, X, Y, \delta, \bar\alpha$.  
Now let us see what happens when we change keys, noting how the new choice for $F_\sharp$ fits into the scheme. Recall that in the just system a change of key from $C$ to $G$ worked out, but at the next level with the key of $D$ one encounters an obstacle. 
The $G$-major scale in the complex scheme is
\begin{eqnarray}
G = \tfrac{3}{4}, \, \,A = \tfrac{5}{6}, \, \,B = \tfrac{2}{3}\xi, \, \,C = 1, \, \,D = \gamma, \, \,E = \tfrac{5}{4}, \, \,F_{\sharp} =  \xi, \, \,G'= \tfrac{3}{2},
\end{eqnarray} 
starting at $G_3$. For the corresponding sequence of atoms we have $1: Y, \delta, \bar\alpha, \gamma, \bar\gamma, \delta, \bar\alpha$. 
In particular, one notices that $\delta = \frac{4}{5}\xi$ lifts the $E = \frac{5}{4}$ up to $F_ {\sharp} =  \xi$.
Then one sees that $\bar \alpha = \frac{3}{4}\bar \xi$ lifts the $F_ {\sharp}$ up to  $G'= \frac{3}{2}$, by use of the relation 
$\xi \,\bar \xi = 2$. 

In the key of $D$ major, there is an added level of complexity on account of the two sharps and the fact that $D$ itself is a complex tone:
\begin{eqnarray}
D = \gamma, \, \,E = \tfrac{5}{4}, \, \,F_ {\sharp} =  \xi, \, \,G= \tfrac{3}{2}, \, \,A = \tfrac{5}{3}, \, \,B = \tfrac{4}{3}\xi, \, \,C'_ {\sharp} = \tfrac{3}{2} \xi, \, \,D' = 2\gamma.
\end{eqnarray} 
For the corresponding sequence of atoms we obtain $1: \bar\gamma,  \delta, \bar\alpha, Y, \delta, X, \beta$.
The treatment of the $F_ {\sharp}$ is like that of the $G$-major scale. 
For the $C_ {\sharp}$, we see that
$X = \frac{9}{8}$  lifts $B = \frac{4}{3}\xi$ to $C_ {\sharp} = \frac{3}{2} \xi$, and then $\beta$ lifts $C_ {\sharp}$ to $D' = 2\gamma$, by use of the identity $\alpha \beta = \gamma$.
Essentially the same line of reasoning can be applied to the keys
of $A$ major, $E$ major and $B$ major. 

The flat keys show a slightly different pattern. For example, in the case of $F$ major, starting at $F_4$, we have 
\begin{eqnarray}
F = \tfrac{4}{3}, \, \,G = \tfrac{3}{2}, \, \,A = \tfrac{5}{3}, \, \,B_ {\flat}  = \tfrac{5}{4}\xi, \, \,C' = 2, \, \,D' = 2 \gamma, \, \,E' = \tfrac{5}{2}, \, \,F' = \tfrac{8}{3},
\end{eqnarray} 
with the aggregation sequence $1: X, Y, \alpha, \bar \delta, \gamma, \bar \gamma, Z$.
Thus the semitone $\alpha = \frac{3}{4} \xi$ lifts $A = \frac{5}{3}$ to $B_ {\flat} = \frac{5}{4} \xi$.
Then $\bar \delta = \frac{4}{5}\bar\xi$ lifts $B_ {\flat}$ to $C' = 2$. 
Similar constructions apply to the keys of $B_\flat$, $E_\flat$,  $A_\flat$, $D_\flat$ and $G_\flat$.

Let us take stock of the complex whole tones and semitones at our disposal. We have the three semitones 
\begin{eqnarray}
\alpha = \tfrac{3}{4}(1+\ri), \quad \beta = \tfrac{1}{3}(3 + \ri), \quad Z=\tfrac{16}{15}, 
\end{eqnarray} 
\noindent along with the four whole tones 
\begin{eqnarray}
X=\tfrac{9}{8}, \quad Y=\tfrac{10}{9}, \quad \gamma =  \tfrac{1}{2} (1 + 2\ri), \quad \delta = \tfrac{4}{5}(1+\ri). 
\end{eqnarray} 
\noindent These intervals can be used to construct complex analogues of the major and minor scales in all twelve keys. The resulting scheme of intervals for the twelve major scales is as follows:\\
\\ \vspace{0.30 cm}
\,\,$C$ major \quad \,\,$1:  \gamma, \bar\gamma, Z, X, Y, \delta, \bar\alpha$ 
\quad \quad \,$F$ major \quad \,\, $\frac{4}{3}:  X, Y, \alpha, \bar\delta, \gamma,\bar\gamma, Z$
\\ \vspace{0.3 cm}
\,\,$G$ major \quad \,\,$\frac{3}{2}:  Y, \delta, \bar\alpha,  \gamma, \bar\gamma,  \delta, \bar\alpha$
\quad \quad \, $B_\flat$ major \quad $\frac{5}{8}\xi:  \bar\delta, \gamma, \bar\beta, \bar\delta, X, Y, \alpha$
\\ \vspace{0.3 cm} 
\,\,$D$ major \quad \,\,$\gamma:  \bar\gamma, \delta, \bar\alpha, Y, \delta, X,\beta $
\,\quad \quad $E_\flat$ major \quad $\frac{5}{6}\xi:  \bar\delta, X, \bar\beta, \gamma, \bar\delta, \gamma, \bar\beta$
\\ \vspace{0.3 cm}
\,\,$A$ major \quad \,\,$\frac{5}{6}:  \delta, X, \beta, \bar\gamma, \delta, \bar\gamma, \beta$
\,\,\quad \quad $A_\flat$ major \quad $\frac{3}{4}\bar\beta: \gamma, \bar\delta, \alpha, Y, \bar\delta, X, \bar\beta$
\\ \vspace{0.3 cm}
\,\,$E$ major \quad \,\,$\frac{5}{4}:  \delta, \bar\gamma, \beta, \delta, X, Y, \bar\alpha$
\,\quad \quad $D_\flat$ major \quad$\frac{3}{4}\xi: Y, \bar\delta, \alpha,  \bar\gamma, \gamma, \bar\delta, \alpha$
\\ \vspace{0.3 cm}
\,$B$ major \quad $\frac{2}{3}\xi:  X, Y, \bar\alpha, \delta,  \bar\gamma, \gamma, Z$
\quad \quad $G_\flat$ major \quad \, $\xi: \bar\gamma, \gamma, Z, X, Y, \bar\delta, \alpha$.
\\ 

\noindent Here, the initial note is given for each major scale, followed by the aggregation sequence that generates that scale. 
 The notes $C$, $E$, $F$, $G$ and $A$ take the Ptolemaic values and for $F_{\sharp}$ we have $\xi = 1 + \rm i$. Each minor scale is the relative minor of an associated major scale.  For example, for $C$ minor we have the sequence $1: \gamma, \bar \beta, \bar \delta, X, \bar \beta, \gamma, \bar \delta$ obtained by permuting the $E_\flat$-major sequence. In this way we obtain a system of major and minor scales in all keys. 
 
%
\section{Well-Tempered Complexity}
\noindent Motivated by the constructions of the previous section, we are now in a position to present the complex scheme in its entirety. We begin by recalling a few results concerning so-called  {\em Gaussian integers}. We write $\mathbb Z$ for the ring of ordinary real integers, which we call {\em rational integers}. We write  $\mathbb Z\,[\,\ri\,]$ for the ring of Gaussian integers, by which we mean complex numbers of the form $a + b \ri$, where $a,b \in \mathbb Z$ and $ i = \sqrt{-1}$. Gaussian integers can be added, subtracted and multiplied to yield other Gaussian integers.
In what follows we often refer to Gaussian integers simply as ``integers". The complex conjugate of an integer $c = a + \ri b$ is denoted $\bar c = a - b i$, and we write $|c| = \sqrt{a^2 + b^2}$ for the magnitude of $c$. 

By a {\em unit} we mean an integer that divides any integer. There are four units: $\pm1, \pm\ri $, and for any integer $n$ we refer to the four numbers $\pm n, \pm \ri n $ as its {\em associates}. By a prime we mean any integer $p$, not $0$, and not a unit, that is divisible only by its associates 
$\pm p, \pm \ri p $ and the units $\pm 1, \pm \ri$. Every integer $n$ can be expressed in the form $n = \Pi_{j=1}^r \,p_j$, where the $p_j$ are primes. 
The fundamental theorem of arithmetic for Gaussian integers asserts that this expression is unique, apart from the order of the primes, the presence of units, and ambiguities between associated primes (Hardy and Wright 1938). 

By a {\em real} prime, we mean a Gaussian prime, in the sense just defined, that is real; the conventional primes are all {\em rational} primes. For example, $3$ and $7$ are real primes, whereas $2 = (1+\ri)(1-\ri)$ and $5= (1+2\ri)(1-2\ri)$ are not; but $2$, $3$, $5$ and $7$ are rational primes. A rational prime $p > 0$ is a real prime only if it is of the form $p = 4n + 3$ for $n \in \mathbb N$.  

A Gaussian integer $n = a + b \ri$, $a,b \in \mathbb Z$, is said to be {\em even} if $a,b$ are both even or both odd in the usual sense, and otherwise $n$ is {\em odd}. For real $n$ this definition of even and odd coincides with the usual one. Then one can show that an integer is even iff it is divisible by $1+\ri$. This generalizes the idea that a rational integer is even iff it is divisible by two.

As usual, we write $\mathbb Q$ for rational numbers, by which we mean numbers of the form $a/b$ for $a,b \in \mathbb Z$ and $b\neq 0$. 

The aggregate of numbers of the form $r + \ri s$, with $r,s \in \mathbb Q$, constitute the so-called Gaussian rationals $\mathbb Q\,[\,\ri\,]$. 

Given the fundamental theorem of arithmetic for Gaussian integers, one can show that any Gaussian rational has a unique factorization of the form $z = \Pi_{j=1}^r \,(p_j)^{a_j}$, where the primes are distinct and the $a_j$ are non-vanishing integers. The uniqueness is qualified up to permutation, the presence of units, and ambiguities between associated primes.

It will have been observed that the tones (or intervals) that we have considered so far, apart from the equal temperament semitones, are Gaussian rationals and thus inherit from $\mathbb Q\,[\,\ri\,]$ the multiplicative group structure that is characteristic of the concatenation of intervals. In practice, we are concerned with Gaussian rationals that can be built up from a small basis of Gaussian primes. 

\begin{Definition} We say that a tone $t$ is {\em complex five-limit} and write 
$t \in \mathcal L(5, \mathbb C)$ if it takes the form 
\begin{eqnarray} \label{five limit}
t = (1 + \ri)^a \, (1 - \ri)^b \, 3^c\, (1 + 2\ri)^d\, (1 - 2\ri)^e  
\end{eqnarray}
for some collection of integers $a,b,c,d,e \in \mathbb Z$. 
\end{Definition}

It should be plain that a real five-limit tone $t \in \mathcal L(5, \mathbb R)$, by which we mean a tone of the form
$t = 2^p \, 3^q \, 5^r$,  
for some choice of $p,q,r \in \mathbb Z$, is necessarily a complex five-limit tone; for clearly, if $t$ is to be real in \eqref{five limit}, we must have $a=b$ and $d = e$. 

More generally, we write $t \in \mathcal L(n, \mathbb C)$ for $n \in \mathbb N$ if $t$ can be expressed as a product of powers of prime factors of rational integers that are of magnitude less than or equal to $n$. Thus, for example, $t \in \mathcal L(3, \mathbb C)$ consists of intervals of the form
\begin{eqnarray}\label{three limit}
t = (1 + \ri)^a \, (1 - \ri)^b \, 3^c.
\end{eqnarray}
It should be evident that $\mathcal L(n, \mathbb C)$ has for each $n \in \mathbb N$ the structure of a free Abelian group and that $\mathcal L(m, \mathbb C)$ is a subgroup of $\mathcal L(n, \mathbb C)$ for $m\leq n$. 

This leads us to comment on the significance of units in our tonal language. The discussion above suggests that if $t \in \mathcal L(5, \mathbb C)$ represents a complex tone and $\epsilon$ is a unit, then $\epsilon t$ represents the same tone as $t$ . Thus, $t$ represents the same musical value as its associates. In particular, $\xi$, $-\xi$, $\ri \xi$, and $-\ri \xi$ all represent the tone $F_\sharp$. But $- \ri \xi = \bar \xi$, so $\xi$ and $\bar \xi$ both represent $F_\sharp$.  

One might wonder then if there is some redundancy in the expression for \eqref{five limit}, since $1+\ri$ and $1-\ri$ are associates, as we have just learned, and hence represent the same prime factor up to a unit. There is indeed some redundancy, but we can use 
\begin{eqnarray}\label{powers of i}
(1 + \ri)^a \, (1 - \ri)^b = (-\ri)^b (1+\ri)^{a+b}
\end{eqnarray}
to put \eqref{three limit} in the form 
\begin{eqnarray} \label{three limit revised}
t = \epsilon \, (1 + \ri)^a  \, 3^b,
\end{eqnarray}
and to put \eqref{five limit} in the form
\begin{eqnarray} \label{five limit revised}
t = \epsilon \,(1 + \ri)^a \, 3^b\, (1 + 2\ri)^c\, (1 - 2\ri)^d ,
\end{eqnarray}
where $\epsilon$ is a unit, since, depending on whether $b = 0, 1, 2, 3$ mod $4$  in \eqref{powers of i} we find that $(-\ri)^b = 1, -\ri, -1, \, \ri$. Then, relabelling the exponents, we obtain \eqref{three limit revised} and  \eqref{five limit revised}. 
Depending on the context, we can use one or the other of the equivalent expressions \eqref{five limit} and  \eqref{five limit revised} as a canonical representation for a given tone in the complex $5$-limit case. 

The latter arises more naturally in prime number theory; but the former arises more directly in music theory. 
In fact, we can take matters further, and apply a similar line of reasoning to the complex prime factors of the number five. As we have pointed out, since $\tfrac{1}{2}(1 + 2\ri) \times \tfrac{1}{2}(1 - 2\ri) = \tfrac{5}{4}$, we can regard each of the complex numbers  $\gamma = \tfrac{1}{2}(1 + 2\ri)$ and $\bar \gamma = \tfrac{1}{2}(1 - 2\ri)$ as representing a type of whole tone interval.  But their quotient is given by the complex phase factor
\begin{eqnarray}
\omega =   \frac{1 - 2\, \rm{i}} {1 + 2\, \rm{i}}   
=  -  \left( \frac{3 + 4\, \rm{i}} {5}  \right).
\end{eqnarray}
We see that  $\omega \, \bar \omega = 1$ and $\bar \gamma = \omega \gamma$. Thus, if we assume that  $\gamma$ and $\bar \gamma$ represent the same tone, one can ask what the situation is if higher powers of $\omega$ are brought into play. 

\begin{Proposition}
A complex tone $\zeta$ will be said to be a phase factor if $\zeta \,\bar \zeta = 1$. A phase factor is complex $5$-limit if and only if for some unit $\epsilon$ and some $k \in \mathbb Z$ it is of the form
\begin{eqnarray}\label{zeta}
\zeta = \epsilon \left( \frac{3 + 4\, \rm{i}} {5}  \right)^k.
\end{eqnarray}
\end{Proposition}

\noindent \textit{Proof}. If $\zeta$ is a complex $5$-limit tone then it can be put into the form \eqref{five limit}. Then for it to be a phase factor we require $b = -a$, $c = 0$ and $e = -d$. This implies that 
\begin{eqnarray}\label{expression for zeta}
\zeta = \left( \frac{1 -\, \rm{i}} {1 +\, \rm{i}}  \right)^j \left( \frac{1 - 2\, \rm{i}} {1 + 2\, \rm{i}}   \right)^k
=  (-\ri)^j \, (-1)^k   \left( \frac{3 + 4\, \rm{i}} {5}  \right)^k
\end{eqnarray}
for some $j,k \in \mathbb Z$,
which shows that any complex $5$-limit phase factor is of the form \eqref{zeta}. Conversely,  any tone of the form \eqref{zeta} is indeed a complex $5$-limit phase factor. \qed

\vspace{.25cm}

\noindent {\bf Remark 3}. Thus, by virtue of \eqref{five limit} and \eqref{expression for zeta} it follows that any complex $5$-limit tone admits a unique factorization of the form 
\begin{eqnarray}
t =  (1 + \ri)^a \, 3^b\, (1 + 2\ri)^c ,
\end{eqnarray}
modulo units and complex $5$-limit phase factors. 

\vspace{.25cm}

\noindent {\bf Remark 4}.  The space $\mathcal L(5, \mathbb C)$, modulo such factors, forms a Generalized Interval System (GIS) in the sense of Lewin (1987). We have a triple $(S, I, \rm{int})$, where $S = \mathcal L(5, \mathbb C)$ modulo units and complex $5$-limit phases is regarded as a space of complex tones, $I$ is another copy of $\mathcal L(5, \mathbb C)$, modulo such factors, now regarded as a group of intervals, and $\rm{int}$ is the map that takes the complex tones $s, t \in S$ to the complex interval $t/s \in I$, mod units and complex $5$-limit phase factors. 
%
%
\section{Complex Chromatic Scales}
\noindent Remarkably, we can construct a single complex chromatic scale that can be used for the formation of all of the major scales given in the previous section, along with their modal variations. 
This scale is as follows: 
\begin{eqnarray} \label{Scheme I chromatic scale}
&&
C = 1, \, \,
C_{\sharp} =  \tfrac{3}{4} (1 + \ri)  \, \,
D = \tfrac{1}{2} (1+2\ri), \, \,
D_ {\sharp}=  \tfrac{5}{6}(1 + \ri), \, \,
E = \tfrac{5}{4},  \, \,
F =\tfrac{4}{3},  \, \,
F_ {\sharp} =1 + \ri,  \, \,
\nonumber \\ && \quad  \quad
G = \tfrac{3}{2},  \, \,
 G_ {\sharp} = \tfrac{1}{2}(3 - \ri),  \, \,
A =  \tfrac{5}{3},  \, \,
A_ {\sharp} = \tfrac{5}{4}(1 + \ri),  \, \,
B = \tfrac{4}{3}(1 + \ri),  \, \,
C' =  2.
\end{eqnarray} 
\noindent Moreover, this complex chromatic scale can be constructed by aggregation in a very simple way. 
Only three semitones are required and the aggregation sequence starting at $C=1$ is
\begin{eqnarray}  \label{Scheme I aggregating sequence}
1: \alpha, \beta, \bar\beta, \bar \alpha, Z, \alpha,  \bar \alpha, \bar\beta, \beta, \alpha, Z, \bar \alpha,
\end{eqnarray}
where $\alpha = \frac{3}{4}(1+\rm i)$, $\beta = \frac{1}{3}(3 + \ri)$, $Z=\frac{16}{15}$.
Note that the first six intervals in the sequence are complex conjugate to the second six. 
Indeed, one can check that 
\begin{eqnarray}
\alpha \beta \bar\beta \bar \alpha Z  \alpha = 1 + \ri, \quad 
\bar \alpha \bar\beta \beta \alpha Z \bar \alpha = 1 - \ri.
 \end{eqnarray} 

Thus, we have demonstrated the existence of a twelve-tone chromatic scale in the Gaussian rationals, constructed by aggregation from a set of three complex semitones 
$\alpha, \beta, Z \in \mathbb Q\,[\,\ri\,]$ such that 
\begin{eqnarray}
C =1, \,\, E = \tfrac{5}{4}, \,\, F = \tfrac{4}{3}, \,\, F_{\sharp}= 1 + \ri, \,\, G =  \tfrac{3}{2}, \,\, A = \tfrac{5}{3}, \,\, C' = 2, 
 \end{eqnarray} 
 
\noindent and one can check that the chromatic scale \eqref{Scheme I chromatic scale} is consistent with the scheme of major and minor scales introduced in Section \ref{section V}. We are then led to ask if any other complex tuning systems exist based on three semitones. The answer is yes, and surprisingly we find that exactly three such systems exist.  

\begin{Theorem}
Let $S$ denote a set of three complex five-limit semitones satisfying  {\rm (a)} there exists a chromatic scale of twelve notes constructed by aggregation from the elements of $S$ and their complex conjugates such that $C =1$, $E = \frac{5}{4}$, $F = \frac{4}{3}$, $F_{\sharp}= 1 + \ri$, $G =  \frac{3}{2}$, $A = \frac{5}{3}$, $C'=2$, and {\rm (b)} the aggregation is conjugate symmetric, in the sense that the first six elements of the chromatic series are conjugates of the second six elements, in the same order. Then three and only three such sets exist, given by

\vspace{.25cm}

{\rm Complex System I:} $S = \{\tfrac{3}{4} (1 + \ri), \, \tfrac{1}{3} (3 + \ri) \zeta, \, \tfrac{16}{15} \}$, 
$\zeta = \epsilon \left( \frac{3 + 4\, \rm{i}} {5}  \right)^k$, $\epsilon \in \{1, -1, \ri, -\ri\}$, $k \in \mathbb Z$,

\vspace{.25cm}
{\rm Complex System II:} $S = \{\tfrac{3}{4} (1 + \ri), \, \tfrac{20}{27}(1 - \ri), \, \tfrac{16}{15}\}$, 

\vspace{.25cm}
{\rm Complex System III:} $S = \{\tfrac{3}{4} (1 + \ri), \, \tfrac{25}{24}, \,\tfrac{16}{15}\}$.
\end{Theorem}

\noindent \textit{Proof}. Let the sequence of aggregation take the form 
\begin{eqnarray} \label{twelve-tone}
1 : X_1,\,X_2,\,X_3,\,X_4,\,X_5,\,X_6,\,X_7,\,X_8,\,X_9,\,X_{10},\,X_{11},\,X_{12},
 \end{eqnarray} 
 where each of the $X$s
  belong to $S$. Conjugate symmetry implies the sequence takes the form  
\begin{eqnarray} \label{conjugate}
1 : X_1,\,X_2,\,X_3,\,X_4,\,X_5,\,X_6,\,\bar X_1,\,\bar X_2,\,\bar X_3,\,\bar X_{4},\,\bar X_{5},\,\bar X_{6}.
 \end{eqnarray} 
Then we impose the following constraints to fix the designated notes: 
\begin{eqnarray}
X_1\,X_2 \,X_3\,X_4 = \tfrac{5}{4}, \quad X_1\,X_2\,X_3\,X_4\,X_5 =  \tfrac{4}{3}, \quad X_1\,X_2\,X_3\,X_4\,X_5\,X_6 = 1+\ri, \nonumber
 \end{eqnarray} 
 \vspace{- 0.75cm}
\begin{eqnarray}
X_1\,X_2\,X_3\,X_4\,X_5\,X_6\,\bar X_1 = \tfrac{3}{2}, \quad X_1\,X_2\,X_3\,X_4\,X_5\,X_6\,\bar X_1\,\bar X_2\,\bar X_3 = \tfrac{5}{3}.
 \end{eqnarray} 
These relations give $X_1 = \alpha$, $X_2\, X_3 = \tfrac{10}{9}$, $X_4 = \bar  \alpha$, $X_5 = Z$, and $X_6 = \alpha$, 
where $\alpha = \tfrac{3}{4} (1+\ri)$ and $Z = \tfrac{16}{15}$. Thus, $S$ must contain $\alpha$ and $Z$ alongside another particle. This remaining particle is determined by the constraint $X_2\, X_3 = \tfrac{10}{9}$, and there are three possibilities: 

 \underline{Complex System I}. $X_2$ and $X_3$ are both new particles in this case, and hence complex conjugates, since there can be no more than three particles altogether. Thus we set $X_2 = \beta$ and $X_3 = \bar \beta$ for some new semitone $\beta$. The relation $\beta \, \bar \beta = \tfrac{10}{9}$ then implies
\begin{eqnarray}
\beta = \tfrac{1}{3} (3 + \ri) \zeta,
\end{eqnarray}
for some phase factor $\zeta$. Imposing the condition that $\beta$ should be complex five-limit and using Proposition 1, one finds the solution for $\beta$ is by
\begin{eqnarray}
\beta = \tfrac{1}{3} (3 + \ri) \, \epsilon \left( \frac{3 + 4\, \rm{i}} {5}  \right)^k,
\end{eqnarray}
where $\epsilon$ is a unit and $k$ is an integer. 
Thus for System I we obtain
\begin{eqnarray}
S = \{ \alpha, \beta, Z\}. 
\end{eqnarray}
This is identical to the example we considered earlier in this section, only now with the inclusion of a phase factor parametrized by $\epsilon$ and $k$. If tones correspond to equivalence classes of elements of $\mathcal L(5, \mathbb C)$ modulo units and phase factors, then $S$ is uniquely determined. 

\underline {Complex System II}. In this case one of the semitones $X_2$ and $X_3$ is taken to be $\alpha$ or $\bar \alpha$. Let us say $X_2 = \bar \alpha$. Then $X_2\, X_3 = \tfrac{10}{9}$ implies $X_3 = \frac{20}{27}(1+\ri)$. But if we express $X_3$ as a multiple of $ \alpha$, then we obtain $X_3 = \kappa^{-1} \, \alpha$,
and we see that the syntonic comma makes an appearance. Thus for System II we have
\begin{eqnarray}
S = \{ \alpha, \alpha', Z\},
\end{eqnarray}
where $\alpha' = \kappa^{-1} \,\alpha$. Note that the two complex semitones are proportional in this case, with a real factor of proportionality. 
The relation between $\alpha$ and $\alpha'$ is analogous to that between $X$ and $Y$ in the just system, where $Y = \kappa^{-1} \, X$.  Note that instead of $X_2 = \bar \alpha$,  we could have taken  $X_2 =  \alpha$, or $X_3 = \alpha$, or $X_3 = \bar \alpha$. Thus, System II occurs in four variants. But if tones are equivalence classes of elements of $\mathcal L(5, \mathbb C)$ modulo units and complex $5$-limit phase factors, then there are only two variants, since $\alpha$ and $\bar\alpha$ are equivalent modulo units.

\underline {Complex System III}. Here one of the semitones $X_2$ and $X_3$ is taken to be $\tfrac{16}{15}$. Let us say $X_3 = \tfrac{16}{15}$. Then $\tfrac{16}{15}\,X_2  = \tfrac{10}{9}$ which gives 
$X_2 = \tfrac{25}{24}$. Thus for System III we have
\begin{eqnarray}
S = \{ \alpha, \rho, Z\},
\end{eqnarray}
where $\rho = \tfrac{25}{24}$. The interval $\rho$ is well known in the theory of tuning and is variously referred to as the minor chromatic semitone or the small half-tone. Again, we could have taken $X_2 = \tfrac{16}{15}$ and $X_3 = \rho$, so we see that System III occurs in two variants.
\qed 
\vspace{0.5cm}

We have already presented the complex chromatic scale and the associated aggregation sequence for System I at \eqref{Scheme I chromatic scale} and \eqref{Scheme I aggregating sequence}. These values are recorded in Table I  
along with the details of the other two systems of complex tonalities with three semitones. Under System II, a calculation shows that the aggregation sequence starting at $C = 1$ is
\begin{eqnarray}
 1: \alpha,\, \bar \alpha,\, \alpha',\, \bar \alpha, \, Z, \, \alpha,\, \bar \alpha, \,\alpha,\, \bar \alpha',\, \alpha, \,Z, \, \bar \alpha
\end{eqnarray}
and the chromatic scale is given by
\begin{eqnarray}
&&
C = 1, \, \,
C_{\sharp} =  \tfrac{3}{4} (1 + \ri),  \, \,
D = \tfrac{9}{8}, \, \,
D_ {\sharp}=  \tfrac{5}{6}(1 + \ri), \, \,
E = \tfrac{5}{4},  \, \,
F =\tfrac{4}{3},  \, \,
F_ {\sharp} =1 + \ri,  \, \,
\nonumber \\ && \quad  
G = \tfrac{3}{2},  \, \,
 G_ {\sharp} = \tfrac{9}{8}(1 + \ri),  \, \,
A =  \tfrac{5}{3},  \, \,
A_ {\sharp} = \tfrac{5}{4}(1 + \ri),  \, \,
B = \tfrac{4}{3}(1 + \ri),  \, \,
C' =  2.
\end{eqnarray} 

\noindent This scale is characterized by the fact that the notes $C$, $D$, $E$, $F$, $G$, $A$, $C'$ take on just values, whereas the remaining notes are given by multiples of $\xi$. In the variant shown, we have taken $X_2 = \bar \alpha, X_3 = \alpha'$, which leads to $D = \tfrac{9}{8}$. In the variant where $X_2 = \bar\alpha', X_3 =  \alpha$, we obtain $D = \tfrac{10}{9}$. In both cases there is a wolf interval, either between $D$ and $A$ or between $D$ and $G$. 
See Table II for further details of the variants of System II. 
Turning now to System III, we see that under its first variant the aggregation sequence takes the form
\begin{eqnarray}
1: \alpha,\, Z,\, \rho,\, \bar \alpha,\, Z,\, \alpha,\,  \bar \alpha,\, Z, \,\rho,\, \alpha,\, Z,\, \bar \alpha
\end{eqnarray}
and for the chromatic scale we obtain

\begin{eqnarray}
&&
C = 1, \, \,
C_{\sharp} =  \tfrac{3}{4} (1 + \ri),  \, \,
D = \tfrac{4}{5} (1 + \ri), \, \,
D_ {\sharp}=  \tfrac{5}{6}(1 + \ri), \, \,
E = \tfrac{5}{4},  \, \,
F =\tfrac{4}{3},  \, \,
F_ {\sharp} =1 + \ri,  \, \,
\nonumber \\ && \quad  \quad \quad \,\,\,\,
G = \tfrac{3}{2},  \, \,
 G_ {\sharp} = \tfrac{8}{5},  \, \,
A =  \tfrac{5}{3},  \, \,
A_ {\sharp} = \tfrac{5}{4}(1 + \ri),  \, \,
B = \tfrac{4}{3}(1 + \ri),  \, \,
C' =  2.
\end{eqnarray}

\noindent Note that we have the simple ratio $A/D = \rho \, \xi$. We also have $G_{\sharp} = \tfrac{8}{5}$, which is the just value of this note. In the second variant of System III, we have $D =  \tfrac{25}{32} \xi$ and $G_{\sharp} = \tfrac{25}{16}$, the latter being the so-called augmented fifth, which can be thought of as a compounding of two intervals of a third. In both variants of System III one has $G_{\sharp} /D = \bar \xi$.  See Table III. 

\section{Complex Pythagorean Systems}

\noindent It is not unreasonable to ask whether one can construct complex chromatic systems based on two semitones, rather than three. Whereas the point of departure for the three-semitone complex system was  the just scale, the case of two complex semitones begins with consideration of the ancient Pythagorean system, the elements of which we briefly recall. 

We use the term ``Pythagorean" to refer broadly to the entire class of tuning systems for which intervals are products of powers of the numbers two and three (Thirring 1990, Halu$\check{\rm{s}}$ka 1999).
If we begin at $C=1$, then an ascending sequence of tones for $C, G, D, A, . \,.\,.\,$ can be constructed using the cycle of fifths by multiplying successive terms by $\frac{3}{2}$ and then normalizing with a power of two to bring the pitch back into the original octave. We obtain 
\begin{eqnarray}
&&
C = 1, \, \,
G = \tfrac{3}{2},  \, \,
D = \tfrac{9}{8}, \, \,
A = \tfrac{27}{16},  \, \,
E = \tfrac{81}{64},  \, \,
B = \tfrac{243}{128},  \, \,
F_ {\sharp} = \tfrac{729}{512}, \, .\,\, .\, \,.\,\, ,
\end{eqnarray} 
and so on up the cycle of fifths, but lowering the tones where necessary by one or more octaves to ensure that they all fall within the same octave. A descending sequence can be similarly constructed by multiplying successive terms by $\frac{2}{3}$, and normalizing, to give
\begin{eqnarray}
C = 1, \, \,
F = \tfrac{4}{3},  \, \,
B_ {\flat} = \tfrac{16}{9},  \, \,
E_ {\flat}= \tfrac{32}{27}, \, \,
A_ {\flat} = \tfrac{128}{81},  \, \,
D_{\flat} = \tfrac{256}{243}, \, \,
G_{\flat} = \tfrac{1024}{729},\, .\,\, .\,\,.\,\, ,
\end{eqnarray} 
and so on down the cycle of fifths. The resulting $F_ {\sharp}$ is different from the $G_{\flat}$, but if we choose one or the other of them then we merge the two series to obtain a Pythagorean version of the chromatic scale, given (in this case, with the $G_ {\flat}$) by
\begin{eqnarray} \label{Pythagorean chromatic scale}
&&
C = 1, \, \,
D_{\flat} = \tfrac{256}{243}, \, \,
D = \tfrac{9}{8}, \, \,
E_ {\flat}= \tfrac{32}{27}, \, \,
E = \tfrac{81}{64},  \, \,
F = \tfrac{4}{3},  \, \,
G_{\flat} = \tfrac{1024}{729}, \, \,
\nonumber \\ && \quad \, \, G = \tfrac{3}{2},  \, \,
A_ {\flat} = \tfrac{128}{81},  \, \,
A = \tfrac{27}{16},  \, \,
B_ {\flat} = \tfrac{16}{9},  \, \,
B = \tfrac{243}{128},  \, \,
C' = 2.
\end{eqnarray} 
It is straightforward to verify that this scale is generated by aggregation from a pair of semitones, given by 
\begin{eqnarray}
\sigma = \tfrac{256}{243}, \quad \tau = \tfrac{2187}{2048}, 
\end{eqnarray} 
corresponding respectively to the $D_{\flat}$ of the descending series and the $C_{\sharp}$ of the ascending series (one step beyond the $F_ {\sharp}$ that we have shown). The aggregation sequence is 
\begin{eqnarray}
1: \sigma, \,\tau, \,\sigma,\, \tau,\, \sigma,\, \sigma,\,  \tau, \, \sigma, \,\tau,\, \sigma, \,\tau,\, \sigma.
\end{eqnarray}
The fact that the {\em diesis} $\sigma$ appears seven times and {\em apotome} $\tau$ appears five times is reflected in the identity $\sigma^7 \tau^5 = 2$, which is easily checked by use of the
relations $\sigma = 2^8 \, 3^{-5}$ and $\tau = 2^{-11} \, 3^7$. See Table IV.  The origins of the Pythagorean system are shrouded in mystery, but it should be evident that both the {\em pentatonic} scale, 
\begin{eqnarray}
&&
C = 1, \, \,
D = \tfrac{9}{8}, \, \,
F = \tfrac{4}{3},  \, \,
G = \tfrac{3}{2},  \, \,
A = \tfrac{27}{16},  \, \,
C' = 2,
\end{eqnarray} 
and the {\em heptatonic} scale
\begin{eqnarray}
&&
C = 1, \, \,
D = \tfrac{9}{8}, \, \,
E = \tfrac{81}{64},  \, \,
F = \tfrac{4}{3},  \, \,
G = \tfrac{3}{2},  \, \,
A = \tfrac{27}{16},  \, \,
B = \tfrac{243}{128},  \, \,
C' = 2
\end{eqnarray} 
can be generated aggregatively within this system by pairs of certain atoms, where each atom has a certain particle composition. 

In the case of the pentatonic scale, we have the whole tone $X = \frac{9}{8}$ given by $\sigma \tau$ and the minor third $W = \frac{32}{27}$ given by  $\sigma^2 \tau$. The associated aggregation sequence starting at $C=1$ is $1: X, W, X, X, W$. 
But in the case of the heptatonic scale, the system is generated by the whole tone $X$ and the semitone $\sigma = \frac{256}{243}$, with the aggregation sequence 
\begin{eqnarray}
1: X,\, X, \,\sigma, \,X,\, X, \,X,\, \sigma.
\end{eqnarray} 

One observes that if we sharpen the notes $E$, $A$ and $B$ of Ptolemy's intense diatonic scale \eqref{just diatonic scale} by applying the syntonic comma, then we obtain the Pythagorean heptatonic scale. In particular, we have
\begin{eqnarray}
\tfrac{5}{4} \xlongrightarrow{\kappa} \tfrac{81}{64},  \quad \quad
\tfrac{5}{3} \xlongrightarrow{\kappa} \tfrac{27}{16}, 
\quad \quad \tfrac{15}{8} \xlongrightarrow{\kappa} \tfrac{243}{128}.
\end{eqnarray} 

Once we have selected a twelve-tone chromatic Pythagorean scale by taking a strip of twelve successive notes from the infinite sequence rising above and descending below $C$, we can for convenience re-write our chromatic scale in such a way that the black keys are represented by notes with sharps, and we have
\begin{eqnarray}
&&
C = 1, \, \,
C_{\sharp} = \tfrac{256}{243}, \, \,
D = \tfrac{9}{8}, \, \,
D_ {\sharp}= \tfrac{32}{27}, \, \,
E = \tfrac{81}{64},  \, \,
F = \tfrac{4}{3},  \, \,
F_ {\sharp} = \tfrac{1024}{729},  \, \,
\nonumber \\ && \quad \, \, G = \tfrac{3}{2},  \, \,
 G_ {\sharp} = \tfrac{128}{81},  \, \,
A = \tfrac{27}{16},  \, \,
A_ {\sharp} = \tfrac{16}{9},  \, \,
B = \tfrac{243}{128},  \, \,
C' = 2.
\end{eqnarray} 
This expression for the chromatic scale is equivalent to that of \eqref{Pythagorean chromatic scale}, but is rather more well suited for a comparison to the scales we have developed based on three semitones. 
As in the case of three semitones, one sees  that the $F_{\sharp}$ is composed of high powers of primes. 

Thus we are prompted to search for a {\it complex} \,Pythagorean system involving lower powers of primes. Following the method of Theorem 2, our approach will be (i) to take as given the real Pythagorean tones of the pentatonic scale, (ii) to supplement these with $F_{\sharp} = 1+ \ri$, and (iii) to impose conjugate symmetry. We obtain the following:

\begin{Theorem}
Let $S$ denote a system of two complex semitones satisfying  {\rm (a)} there exists a chromatic scale of twelve notes constructed by aggregation from elements of $S$ and their complex conjugates such that $C =1$, $D = \frac{9}{8}$, $F = \frac{4}{3}$, $F_{\sharp}= 1+\ri$, $G =  \frac{3}{2}$, $A = \frac{27}{16}$, $C' =2$ and {\rm (b)} the aggregation is conjugate symmetric, in the sense that the first six elements of the chromatic series are conjugates of the second six, in the same order. Then $S = \{ \tfrac{3}{4}(1+\ri),  \tfrac{256}{243}$\}.
\end{Theorem}
\noindent \textit{Proof}. Let the aggregation sequence take the form \eqref{twelve-tone}. When we impose conjugate symmetry the sequence reduces to \eqref{conjugate}, as in the case of three semitones.
Then we impose the following constraints to fix the notes of the pentatonic scale together with $F_{\sharp}= 1+\ri$: 
\begin{eqnarray}
X_1\,X_2  = \tfrac{9}{8}, \quad X_1\,X_2\,X_3\,X_4\,X_5 =  \tfrac{4}{3}, \quad X_1\,X_2\,X_3\,X_4\,X_5\,X_6 = 1+\ri,  \nonumber
 \end{eqnarray} 
 \vspace{- 0.75cm}
\begin{eqnarray}
X_1\,X_2\,X_3\,X_4\,X_5\,X_6\,\bar X_1 = \tfrac{3}{2}, \quad X_1\,X_2\,X_3\,X_4\,X_5\,X_6\,\bar X_1\,\bar X_2\,\bar X_3 = \tfrac{27}{16}.
 \end{eqnarray} 
These relations can be solved to deduce that $X_1 = \alpha$, $X_2 =\bar \alpha$, $X_3 = \alpha$, $X_6 =  \alpha$, and $X_4  X_5 =\tfrac{64}{81} \xi$.
Taking into account that we need to have two semitones in total, we find that the system is uniquely determined to be
$S = \{ \alpha, \sigma\}$,
where $\alpha =  \tfrac{3}{4} (1 + \ri)$ and $\sigma = \tfrac{256}{243}$. 
\qed

\vspace{.5cm}
There are two variants: (i) $X_4 =\bar \alpha$,  
 $X_5 = \tfrac{256}{243}$, and (ii) $X_4 = \tfrac{256}{243}$, 
 $X_5 = \bar \alpha$. The resulting chromatic scale thus takes two different possible forms, depending on the choice of variant. This affects the notes $E$ and $A_{\sharp}$.  For variant (i) we obtain
\begin{eqnarray}
&&
C = 1, \, \,
C_{\sharp} =  \tfrac{3}{4} (1 + \ri)  \, \,
D = \tfrac{9}{8}, \, \,
D_ {\sharp}=  \tfrac{27}{32}(1 + \ri), \, \,
E = \tfrac{81}{64},  \, \,
F =\tfrac{4}{3},  \, \,
F_ {\sharp} =1 + \ri,  \, \,
\nonumber \\ && \, \, \,
G = \tfrac{3}{2},  \, \,
 G_ {\sharp} = \tfrac{9}{8}(1 + \ri),  \, \,
A =  \tfrac{27}{16},  \, \,
A_ {\sharp} = \tfrac{81}{64}(1 + \ri),  \, \,
B = \tfrac{4}{3}(1 + \ri),  \, \,
C' =  2,
\end{eqnarray} 
with the aggregation sequence
\begin{eqnarray}
1: \alpha, \,\bar \alpha,\, \alpha,\,\bar \alpha,\, \sigma,\, \alpha,\, \bar \alpha, \,\alpha,\, \bar \alpha,\, \alpha,\, \sigma,\, \bar \alpha. 
\end{eqnarray} 
Then for variant (ii) we obtain
\begin{eqnarray}
&&
C = 1, \, \,
C_{\sharp} =  \tfrac{3}{4} (1 + \ri)  \, \,
D = \tfrac{9}{8}, \, \,
D_ {\sharp}=  \tfrac{27}{32}(1 + \ri), \, \,
E = \tfrac{8}{9}(1 + \ri),  \, \,
F =\tfrac{4}{3},  \, \,
F_ {\sharp} =1 + \ri,  \, \,
\nonumber \\ && \quad \quad \quad
G = \tfrac{3}{2},  \, \,
 G_ {\sharp} = \tfrac{9}{8}(1 + \ri),  \, \,
A =  \tfrac{27}{16},  \, \,
A_ {\sharp} = \tfrac{16}{9},  \, \,
B = \tfrac{4}{3}(1 + \ri),  \, \,
C' =  2,
\end{eqnarray} 
with the aggregation sequence
\begin{eqnarray}
1: \alpha,\, \bar \alpha,\, \alpha,\, \sigma, \,\bar \alpha,\, \alpha,\, \bar \alpha,\, \alpha, \,\bar \alpha,\, \sigma,\,\alpha,\,  \bar \alpha. 
\end{eqnarray}

\noindent {\bf Remark 5}. It is interesting to note that it was not necessary to assume that the semitones are complex 3-limit in the formulation of Theorem 3. Rather, this feature emerges as a {\it consequence} of the weaker conditions that we have imposed, in contrast with Theorem 2, where it was assumed that the semitones are complex five-limit.

\section{Back to Reality}
\noindent One can ask whether the complex tone systems that we have considered give rise to real ``shadow" systems, where the magnitudes of the complex numbers under consideration might play a role. 
It would thus seem natural to look for an analogue of Theorem 2 in which the role of $\xi$ is replaced by $ |\, \xi \,| = \sqrt 2$. This takes us out of the rational framework, but is nevertheless of interest since the results run in parallel with the complex system in a nice way.  Since one has
$F_{\sharp}= \sqrt{2}$ in equal temperament, as well as in historic tuning systems (for example, that of  Grammateus 1518) where $\sqrt {2}$ arises as the geometric mean between $F = \frac{4}{3}$ and $G =  \frac{3}{2}$ (Barbour 1953), it makes sense to pursue the matter. We obtain the following: 
\begin{Theorem}
Let $S$ denote a system of three real semitones satisfying  {\rm (a)} there exists a chromatic scale of twelve notes constructed by aggregation from elements of $S$ such that $C =1$, $E = \frac{5}{4}$, $F = \frac{4}{3}$, $F_{\sharp}= \sqrt{2}$, $G =  \frac{3}{2}$, $A = \frac{5}{3}$, $C' = 2$, and {\rm (b)} the aggregation is  symmetric, in the sense that the first six elements of the chromatic series equal the second six elements, in the same order. Then there are three such systems, given by

\vspace{.25cm}
{\rm Shadow System I:} $S = \{\tfrac{3}{4} \sqrt{2}, \,\frac{1}{3}\sqrt{10}, \,\tfrac{16}{15}\}$, 

\vspace{.25cm}
{\rm Shadow System II:} $S = \{\tfrac{3}{4} \sqrt{2},\, \tfrac{20}{27}\sqrt{2}, \,\tfrac{16}{15}\}$, 

\vspace{.25cm}
{\rm Shadow System III:} $S = \{\tfrac{3}{4} \sqrt{2}, \,\tfrac{25}{24},\, \tfrac{16}{15}\}$.
\end{Theorem}

\vspace{.2cm}
\noindent \textit{Proof}. 
The aggregation sequence is evidently of the form  
\begin{eqnarray} \label{symmetric system}
1: X_1,\,X_2,\,X_3,\,X_4,\,X_5,\,X_6,\, X_1,\, X_2,\, X_3,\, X_{4},\, X_{5},\, X_{6},
 \end{eqnarray} 
with real entries, and the constraints are given by
\begin{eqnarray}
X_1\,X_2 \,X_3\,X_4 = \tfrac{5}{4}, \quad X_1\,X_2\,X_3\,X_4\,X_5 =  \tfrac{4}{3}, \quad X_1\,X_2\,X_3\,X_4\,X_5\,X_6 =  \sqrt{2}, 
\nonumber
 \end{eqnarray} 
\vspace{- 0.75cm}
\begin{eqnarray}
X_1\,X_2\,X_3\,X_4\,X_5\,X_6\,X_1 = \tfrac{3}{2}, \quad X_1\,X_2\,X_3\,X_4\,X_5\,X_6\,X_1\, X_2\, X_3 = \tfrac{5}{3}.
 \end{eqnarray} 
It follows that $X_1 = X_4 =X_6 =  \theta$, $X_5 =\tfrac{16}{15}$, and $X_2  X_3 =\tfrac{10}{9}$, where $\theta= \frac{3}{4} \sqrt{2}$.
The need for  three semitones in total implies that either (i) $X_2 X_3 = \eta^2$,  or
  (ii) $X_2 X_3 = \theta \, \eta$, or else (iii) $X_2 X_3 = \tfrac{16}{15}\, \eta$,
where $\eta$ is another semitone. Then by solution of the constraint $X_2  X_3 =\tfrac{10}{9}$ we are  led to the three systems in the claimed result. 
\qed

\vspace{.2cm}
As in Theorem 2, the first and third semitones are the same in all three systems, and it is the second semitone that differs. Note that Systems II and III each have two variants, depending on the order of the semitones $X_2$ and $X_3$. 

An analogous result can be obtained for the real shadow of the complex Pythagorean system, where we look at the construction of a real two-semitone system matching the Pythagorean pentatonic scale and satisfying $F_{\sharp} = \sqrt{2}$.

\vspace{.15cm}

\begin{Theorem}
Let $S$ denote a system of two real semitones satisfying  {\rm (a)} there exists a chromatic scale of twelve notes constructed by aggregation from elements of $S$ such that $C =1$, $D = \frac{9}{8}$, $F = \frac{4}{3}$, $F_{\sharp}= \sqrt{2}$, $G =  \frac{3}{2}$, $A = \frac{27}{16}$, $C' =2$ and {\rm (b)} the aggregation is  symmetric in the sense that the first six elements of the chromatic series equal the second six elements in the same order. 
Then there are two such systems, given by

\vspace{.15cm}
{\rm Pythagorean Shadow System I:} $S = \{ \tfrac{3}{4} \sqrt{2}, \,\tfrac{256}{243} \}$, 

\vspace{.15cm}
{\rm Pythagorean Shadow System II:} $S = \{ \tfrac{3}{4} \sqrt{2},\, \frac{8}{9} \sqrt[4]{2}  \}$. 
\end{Theorem}

\vspace{.2cm}
\noindent \textit{Proof}. The aggregation sequence takes the form  \eqref{symmetric system},
with real entries, and the constraints 
\begin{eqnarray}
X_1\,X_2  = \tfrac{9}{8}, \quad X_1\,X_2\,X_3\,X_4\,X_5 =  \tfrac{4}{3}, \quad X_1\,X_2\,X_3\,X_4\,X_5\,X_6 =  \sqrt{2}, 
\nonumber
 \end{eqnarray} 
\vspace{- 0.75cm}
\begin{eqnarray}
X_1\,X_2\,X_3\,X_4\,X_5\,X_6\,X_1 = \tfrac{3}{2}, \quad X_1\,X_2\,X_3\,X_4\,X_5\,X_6\,X_1\, X_2\, X_3 = \tfrac{27}{16}.
 \end{eqnarray} 
It follows that $X_1 = X_2 = X_3 = X_6 =  \frac{3}{4} \,\sqrt{2}$ and $X_4  X_5 =\tfrac{64}{81} \,\sqrt{2}$.
That there should be two semitones implies that either $X_4 \, X_5 =\theta\, \sigma $, where $\theta = \frac{3}{4}\, \sqrt{2}$ and $\sigma = \frac{256}{243}$ is the second semitone, or $X_4  \,X_5 =\nu^2$, where $\nu = \frac{8}{9}\, \sqrt[4]{2} $ is the second semitone, and this gives the systems stated in the theorem. Note that for System I there are two variants: (a) $X_4 =\theta$,  
 $X_5 = \sigma$, (b) $X_4 = \sigma$, $X_5 = \theta$.
\qed

 \vspace{.25cm}
 
{\bf Remark 6}. Pythagorean Shadow System I is based on the magnitudes of the complex Pythagorean system presented in Theorem 3, but  Pythagorean Shadow System II arises independently as a second solution of the conditions of Theorem 5.

\section{Applications, Conjectures, and Conclusions} 

\noindent Complex tonality opens up the possibility of genuinely new types of musical structures and perhaps also helps us to understand existing structures. This becomes evident in the various harmonic relations holding between  scales and between chords. In our Complex System I, all the major scales are different from one another. Thus if we play a simple melody in one key and then transpose it to a different key, the new melody is usually not quite the same as the original. In other words, the various major keys are not transpositionally equivalent.  The same is true of the minor keys. In the key of $C$ major, the tonic and subdominant triads are equivalent (they have the same interval structure) but the tonic and the dominant triads are not, on account of the complex values for $B$ and $D$. Thus for $C$ major and $F$ major we have the triads $[1, \,\frac{5}{4}, \,\frac{3}{2}]$ and $[\frac{4}{3},\, \frac{5}{3},\, 2]$, but for $G$ major we have  the triad $[\frac{3}{2}, \frac{4}{3}\,\xi, \,\frac{8}{5}\,\xi]$. 

The hallmark of the dominant in this case is that the interval $B/G$ is given by  $\frac{8}{9}\,\xi$ and that the interval $D/G$ is given by $\frac{16}{15}\,\xi$, whereas the interval $D/B$ is given by $\frac{6}{5}$, as in the just system. This closeness of relationship between the $F$-major and the $C$-major triads may contribute to the perceived gentleness of the plagal resolution from $F$ major to $C$ major in contrast to the more robust authentic resolution from $G$ major to $C$ major. 

A well-known problem in the theory of harmony is the question of why major and minor chords should be so different in character from one another to the ears. After all, they are composed of the same intervals, $\frac{5}{4}$ and $\frac{6}{5}$, merely stacked in the opposite order. There is no obvious acoustic, physiological or psychological reason why they should produce such a different effect on the listener. 
One might argue along the lines of cultural conditioning, and to be sure, such elements may come in; but surely that cannot be the whole story. In Complex System I,  the $C$-major chord takes the familiar form 
$[1, \,\frac{5}{4}, \,\frac{3}{2}]$, whereas by \eqref{Scheme I chromatic scale} the parallel $C$-minor chord is 
$[1, \,\frac{5}{6} \xi, \,\frac{3}{2}]$, which can be contrasted to the just Aeolian 
$[1, \,\frac{6}{5}, \,\frac{3}{2}]$. But for the relative minor in the complex system, we have  
$[\frac{5}{6},\, 1, \frac{5}{4} ]$, in which the intervals are the same as those of the relative minor in the just system. Thus, in the complex system there is a clear distinction, in the key of $C$ major, between the structure of the  parallel minor (with its complex minor third interval) and the relative minor (with its real minor third interval), whereas in the just system these two minor chords have the same structure. 

Such distinctions  may help us better understand, for example, the startling contrasts in the relations between $C$ major and $C$ minor, on the one hand, and between $C$ major and $A$ minor, on the other hand, in the two movements of Beethoven's Opus 111 Piano Sonata. That the first relation is like that of heaven and hell, and the second like that of acceptance and resignation. One recalls Kretschmar's lecture in Thomas Mann's {\em Doctor Faustus}. 

Our remarks cannot do justice to the profundity of Beethoven's composition; nonetheless, the relations between these two different means of forming a minor chord from a major chord, and the nature of the mental states induced by these chords, are so important that any insight into the way the mind takes on board these distinctions may be of some interest. 

Indeed, if one looks at the Opus 111 Sonata  from the point of view of its embedding in the complex system, more of its structure becomes apparent.
Structural relations between major and minor chords are evident more generally if one views the neo-Riemannian theories (Cohn 1998, 2012) from this angle. For instance, the system of hexatonic relations beginning at $C$ major  involves a progression from the real to the complex and back under Complex System I. The well-known hexatonic cycle
\begin{eqnarray}
C\mbox{ major} \to C\mbox{ minor} \to A_{\flat}\mbox{ major} \to  A_{\flat }\mbox{ minor} \to E\mbox{ major} \to E\mbox{ minor}\to C\mbox{ major}, 
\end{eqnarray} 
\noindent obtained by successive parallel and leading-tone shifts, is given in the Keplerian system by
\begin{eqnarray}
[1, \,\tfrac{5}{4}, \,\tfrac{3}{2}]  \to [1, \,\tfrac{6}{5}, \,\tfrac{3}{2}] \to  [1, \,\tfrac{6}{5}, \,\tfrac{8}{5}]  \to
[\tfrac{15}{16}, \,\tfrac{6}{5}, \,\tfrac{8}{5}] \to [\tfrac{15}{16}, \,\tfrac{5}{4}, \,\tfrac{8}{5}] \to 
[\tfrac{15}{16}, \,\tfrac{5}{4}, \,\tfrac{3}{2}] \to [1, \,\tfrac{5}{4}, \,\tfrac{3}{2}]. 
\end{eqnarray} 
\noindent But in Complex System I this sequence of transformations takes the following form: 
%
%
%
%
%
\begin{eqnarray}
[1, \,\tfrac{5}{4}, \,\tfrac{3}{2} ]  \to 
[1, \,\tfrac{5}{6}(1 + \ri), \,\tfrac{3}{2} ] \to  
[1, \,\tfrac{5}{6}(1 + \ri), \,\tfrac{1}{2}(3 - \ri) ]  \to [\tfrac{2}{3}(1 + \ri), \,\tfrac{5}{6}(1 + \ri), \,\tfrac{1}{2}(3 - \ri) ] \to
\nonumber
\end{eqnarray} 
\vspace{-1cm}
\begin{eqnarray} \hspace{-3cm}
[\tfrac{2}{3}(1 + \ri), \,\tfrac{5}{4}, \,\tfrac{1}{2}(3 - \ri) ] \to 
[\tfrac{2}{3}(1 + \ri), \,\tfrac{5}{4}, \,\tfrac{3}{2} ] \to 
[1, \,\tfrac{5}{4}, \,\tfrac{3}{2}] .
\end{eqnarray} 

\noindent One sees in particular that the starting chord $C$ major has no complex elements, whereas the associated polar chord $A_{\flat }$ minor is constructed entirely from complex elements. Again, more structure is revealed. 

Let us consider another example, once more involving the complex $A_{\flat }$-minor chord. This example concerns the role of the tritone in so-called vagrant chords (Schoenberg 1978).
Ordinarily, the tritone $B/F$ is taken to be a dissonant interval, but there are settings in which it has the opposite effect. 
Such a setting occurs notably, to varying degrees, in the context of the minor sixth chord and its permutations. 
The most well known of these harmonious chords involving the tritone is the $A_{\flat }$-minor sixth, which after permutations we recognize as the Tristan chord, which in just intonation takes the form $[  \frac{2}{3}, \frac{15}{16}, \frac{6}{5}, \frac{8}{5} ]$. 

The remarkable aspect of this chord is that in the presence of the two higher notes the tritone seems to lose its dissonance in such a way that the chord as an \hspace{-.2cm} {\em enti\`eret\'e} acquires that uncanny sense of beauty tinged with tragedy that permeates the opera {\em Tristan und Isolde}. 
We can only speculate (among many others---see, e.g.,~Cohn 2012, Schoenberg 1978) how this comes about, but it may be that in isolation the tritone is perceived by the ear as the dissonant interval $\frac{45}{32}$, with its multiplicity of prime powers; but when nudged by the presence of the other elements of the chord the mind flips (as with optical illusions) to a perception of this interval in its representation by the most basic of the Gaussian primes. 

It is interesting in this context to recall that Lewin (1993) in a notable passage (2.1.5) considers the harmonic intuition associated with the stimulus of the interval $\frac{45}{32}$. In essence he argues (quite persuasively) that the intuition of the interval $F_{\sharp} / C$ takes the $F_{\sharp}$ to be the mediant of the dominant of the dominant, and thus interprets the $F_{\sharp}$ as a leading tone:

\begin{quote} 
\begin{small} Observe that the number $(\cdot \cdot \cdot)$ $\frac{45}{32}$ arises here {\it not} as the product of $2^{-5}$ and $3^2$ and $5$, which is the most natural mathematical factorization. Rather, $\frac{45}{32}$ arises as the product of the four factors $2$, $\frac{5}{4}$, $\frac{3}{4}$ and $\frac{3}{4}$, reflecting its ``natural" way of measuring an intuited chain of intuitions in the given situation.
\end{small}
\end{quote}

\noindent These arguments appear in the context of a wider discussion of the just system, in which musical intervals take the form $2^a \, 3^b \, 5^c$ for $a,b,c \in \mathbb Z$, in connection with which Lewin asserts the following: 

\begin{quote} 
\begin{small} 
It is not immediately clear what {\it intuitions} of ``distance" or ``motion" we are measuring by these intervals. Personally, I am convinced that our intuitions are highly conditioned by cultural factors. In particular, I do not think that the acoustics of harmonically vibrating bodies provide in themselves an adequate basis for grounding those intuitions.
\end{small}
\end{quote}

\noindent Thus it may be that complex intervals map to certain intuitions of chains of intuitions, just as do the intervals of the just system in Lewin's analysis---that an acoustic stimulus may indeed trigger this intuition, in a setting where cultural factors play a significant role. Pursuing the idea further one could then conjecture that 
the setting of the tritone in the environment of the two higher notes of the Tristan chord is sufficient as an acoustic trigger for the interval to map to the complex interval $\xi$ rather, than, say, a leading tone represented by the mediant of the dominant of a dominant---the latter would suggest a resolution into $C$ major, which is clearly not what one observes, despite the many tactics that Wagner does employ in his score to achieve what one might call tentative resolutions. 

In short, we propose that when the tritone is heard in isolation it is registered by the ear as harsh, that when it is heard in the context of, say, an eighteenth century contrapuntal development, it triggers a leading-tone interpretation along the lines of Lewin's gestalt; but that in the setting of the minor sixth chord the mind identifies this interval as the complex prime $1 + \ri$, which it treats as pleasurable, luxurious, even exotic---thus perhaps justifying the term {\em diabolus in musica} which has been applied to this most unstable of intervals.

According to this view, acoustic phenomena can trigger certain intuitions already established in the mind---either innately or through  cultural conditioning---which can then be identified with certain mathematical objects.
 These mathematical objects are constructed from sets of complex 5-limit intervals, which we recognize as melodies, harmonies, rhythms, and so forth. The complex interval system itself can be viewed as an abstraction, and hence the precise form taken by this map from acoustic phenomena to the interval system will inevitably vary somewhat from person to person, from piece to piece, from culture to culture, and over time, even if the existence of such a map can be regarded as a universal. 
 
Under the Complex System I, the Tristan chord takes the form 
$[ \frac{2}{3}, \frac{2}{3}\, \xi, \frac{5}{6}\, \xi, \frac{1}{2} (3 - \rm{i}) ]$. 
Let us put this array under the microscope and deconstruct it into its constituent intervals, of which there are six in all, comparing these to the corresponding intervals arising in the just version of the chord.
For the interval $B$/$F$ we obtain $\xi$ in the complex case, instead of the Keplerian tritone $\frac{45}{32}$, as we have already noted. We take $\xi$ to be consonant since it is a low power of a prime. 
For the interval $E_{\flat}$/$B$ we obtain a perfect third $\frac{5}{4}$ in the complex system, as opposed to the fraction 
$\frac{6}{5} \div \frac{15}{16} = \frac{32}{25}$ that arises in the just system. 
For the dissonant interval $E_{\flat}$/$F$ in the just system we obtain $\frac{9}{5}$. The corresponding interval in the complex system takes the form $\frac{5}{4}\, \xi$, which involves fewer primes (if we do not count powers of 2) and hence can be viewed as no more dissonant that $\frac{9}{5}$.
The interval $A_{\flat}/E_{\flat}$ is given by $\frac{4}{3}$ in the just system. In the complex system, we use the prime factorization $3 - \rm{i} = \xi \, (1 - 2 \rm{i})$ to get 
$A_{\flat}/E_{\flat} = \frac{3}{5} (1 - 2 \rm{i})$, more complicated than the just expression, but  involving low powers of primes, and hence consonant. We can parse this interval as
$A_{\flat}/E_{\flat} = \frac{6}{5}\cdot \frac{1}{2} (1 - 2 \rm{i})$, showing it can be interpreted as the product of a minor third and a complex whole tone. 
This may explain why the fourth between the $E_{\flat}$ and the $A_{\flat}$ is softened, and seems darker in the context of the Tristan chord than is does on its own. The interval $A_{\flat}/F$ is given by $\frac{12}{5}$ in the just system, whereas it takes the form 
$\frac{3}{4}\, \xi \,(1 - 2 \rm{i})$ in the complex system, again with low powers of primes.

Finally, we observe that the sixth between $B$ and $A_{\flat}$ is given by fraction 
$\frac{8}{5} \div \frac{15}{16} = \frac{128}{75}$ in the just system, whereas it takes the form 
$A_{\flat}/B = \frac{3}{4}\,(1 - 2 \rm{i})$ in the complex system, involving only a small number of primes. All the intervals within the Tristan chord arising in the complex system are harmonious (or are at least as harmonious as one might wish), whereas in the just system three of the intervals have large powers of primes and are unacceptable as candidates for consonance. In equal temperament all the intervals are out of tune, though it is difficult to say they are approximately in tune, since one cannot say what form ``exact'' tuning might take. The equally tempered version of the chord sounds acceptable---after all, this is what we hear when we attend the opera or play passages from it on the piano---much as it would have in Wagner's day. It follows that the conventionally performed version of the chord in the setting of the score is sufficient to act as a stimulus for the mind to receive it as a signal for the relevant tones of the complex system, along the lines we have discussed.  We remark, incidentally, that all of the tritone intervals in Complex System I take the value $\xi$, modulo units and 5-limit phase factors. This is consistent with the observation that even under Wagner's many modulations, the Tristan chord does not lose its magic. 

In conclusion, it may be appropriate to make a few brief remarks concerning our research methodology, even if these remarks may be irrelevant to the question of how one might further progress the work. It is not unusual in science for new ideas to arrive in a flash, after lengthy consideration of a body of earlier material, in reaction to which the earlier material can be reorganized and better understood. It may even then look to the outsider as if these developments of the earlier material somehow led to the new ideas, though in reality the sequence of events was opposite. In the present work, our realization that setting $F_{\sharp} = 1 + \rm{i}$ led rather naturally to the construction of complex scales with interesting properties prompted fresh investigations into the foundations of the just system, which in turn led to Theorem 1. 
It is not surprising therefore that our Theorem 1 does not quite have the character of an incremental development based on established ideas in music analysis.  
This is because its methods are fresh and offer new insights, as is evidenced in our analysis of the historical just scales catalogued by Barbour (1953). With Theorem 1 at hand, we were then in a position to investigate the complex system more systematically, leading to Theorem 2. This required bringing in more mathematics, in the form of various results from the theory of Gaussian integers. The connections to word theory,  generalized musical intervals, and the theory of scales came later, and although these latter areas have influenced the presentation of our work, they were not involved in its initial formulation. 

It would thus be disingenuous to suggest that the results we have reported here came about simply as the product of a methodical research program aimed at improving on the just system, somehow eventually leading to the introduction of complex tonality. Nonetheless, going forward, it would make sense to pursue the approach that we have introduced more systematically in the investigation of both real and complex $n$-limit tone systems. That the prime decompositions $2 = (1 + \ri)(1 - \ri)$ and $5 = (1 + 2 \ri)(1 - 2\ri)$ lead to such a wealth of musical structure may come as a surprise. But we can take the view that musical cognition entails the formation of a map from a class of acoustic phenomena to a suitably established tone system, the latter being essentially a mental construct. 
And if the tuning of our ears cannot stretch beyond the confines of 5-limit tones, whether real or complex, then our minds can. In Nietzsche's words, ``Neue Ohren f\"ur neue Musik."  Hence, from the perspective of the mathematician, the complex $n$-limit tone systems may well constitute a natural framework within which music of the future can be composed.

\begin{acknowledgments}
\noindent
We are grateful for comments from participants at RealTime Seminars in the Department of Computing at Goldsmiths University of London, June 2022 and May 2023, and the IMA Conference on Mathematics in Music at the Royal College of Music, London, July 2022, where drafts of this work were presented. We thank J.~Armstrong, T.~Blackwell, D.~M.~Blasius, T.~Crawford, J.~Forth,  G.~Di Graziano, D.~P.~Hewett,  K.~Rietsch and two anonymous referees for helpful comments and suggestions.  
\end{acknowledgments}

\vspace{0.5 cm}
\noindent {\bf References}

\newpage
\vspace{8cm}
\begin{table}
\vspace{3cm}

\begin{tabular}{ |p{.6cm}||p{.4cm}|p{1.5cm}||p{.4cm}|p{1.3cm}||p{.4cm}|p{1.3cm}||  }
 \hline
 \multicolumn{7}{|c|}{\textbf{Complex Tone Systems based on Three Semitones}} \\
 \hline
Note & \multicolumn{2}{c||}{System I} &\multicolumn{2}{c||}{System II}&\multicolumn{2}{c||}{System III}\\
 \hline
 $C$   & 1    &1 &1 & 1&1 & 1\\
 
 $C_{\sharp}$   & $\alpha$    & $\tfrac{3}{4} (1 + \ri)$ & $\alpha$    & $\tfrac{3}{4} (1 + \ri)$ & $\alpha$    & $\tfrac{3}{4} (1 + \ri)$ \\
 
 $D$   & $\beta$    &   $\tfrac{1}{2} (1 + 2\ri)$ & $\bar \alpha$    &   $\tfrac{9}{8}$ & $Z$    &   $\tfrac{4}{5} (1 + \ri)$ \\    
 
 $D_{\sharp}$   & $\bar \beta$   &$\tfrac{5}{6} (1 + \ri)$ & $\alpha'$   &$\tfrac{5}{6} (1 + \ri)$ & $\rho$   &$\tfrac{5}{6} (1 + \ri)$\\
 
 $E$   & $\bar \alpha$    &$\tfrac{5}{4}$ & $\bar \alpha$    &$\tfrac{5}{4}$ & $\bar \alpha$    &$\tfrac{5}{4}$\\ 
 
 $F$   & $Z$    &$\tfrac{4}{3}$  & $Z$    &$\tfrac{4}{3}$ & $Z$    &$\tfrac{4}{3}$ \\
 
 $F_{\sharp}$   & $\alpha$    &$(1+\ri)$ & $\alpha$    &$(1+\ri)$ & $\alpha$    &$(1+\ri)$\\
 
 $G$   & $\bar \alpha$    &$\tfrac{3}{2}$ & $\bar \alpha$    &$\tfrac{3}{2}$  & $\bar \alpha$    &$\tfrac{3}{2}$\\
 
 $G_{\sharp}$   & $\bar \beta$    &$\tfrac{1}{2} (3 - \ri)$ &  $\alpha$    &$\tfrac{9}{8} (1 + \ri)$ &  $Z$    &$\tfrac{8}{5}$ \\ 
 
 $A$   & $\beta$    &$\tfrac{5}{3}$ & $\bar \alpha'$    &$\tfrac{5}{3}$ & $\rho$    &$\tfrac{5}{3}$ \\
 
 $A_{\sharp}$   & $\alpha$    &$\tfrac{5}{4} (1 + \ri)$ & $\alpha$    &$\tfrac{5}{4} (1 + \ri)$ & $\alpha$    &$\tfrac{5}{4} (1 + \ri)$\\ 
 
 $B$   & $Z$    &$\tfrac{4}{3} (1 + \ri)$ & $Z$    &$\tfrac{4}{3} (1 + \ri)$ & $Z$    &$\tfrac{4}{3} (1 + \ri)$\\ 
 
 $C'$ & $\bar \alpha$    &$2$ & $\bar \alpha$    &$2$  & $\bar \alpha$    &$2$\\
  
  \hline
\end{tabular}
\caption{Structures of the three complex tone systems based on three semitones. Each system occurs in a number of variants, one example of which is presented for each system. For each of the three complex systems, the first column gives the aggregation sequence and the second column gives the values taken by the notes of the scale. \vspace{0.25cm} \newline Notation: $\alpha = \tfrac{3}{4} (1 + \ri)$, \,$\alpha' = \tfrac{80}{81} \alpha$, \,$\beta = \tfrac{1}{3} (3 + \ri)$, \,$Z = \frac{16}{15}$, \,$\rho = \frac{25}{24}$.}
\end{table}

\newpage
\begin{table}
\vspace{3cm}
\begin{tabular}{ |p{.6cm}||p{.4cm}|p{1.3cm}||p{.4cm}|p{1.3cm}||p{.4cm}|p{1.4cm}||p{.4cm}|p{1.4cm}||  }
 \hline
 \multicolumn{9}{|c|}{\textbf{Complex System II}} \\
 \hline
Note & \multicolumn{2}{c||}{$X_2 = \bar \alpha, \,X_3 = \alpha'$} &\multicolumn{2}{c||}{$X_2 = \alpha, \, X_3 = \bar\alpha'$}&\multicolumn{2}{c||}{$X_2 = \alpha', \, X_3 = \bar \alpha$}&\multicolumn{2}{c||}{$X_2 = \bar\alpha', \, X_3 = \alpha$}\\
 \hline
 $C$   & 1    &1 &1 & 1&1 & 1 &1 & 1\\
 
 $C_{\sharp}$  & $\alpha$    & $\tfrac{3}{4} (1 + \ri)$ & $\alpha$    & $\tfrac{3}{4} (1 + \ri)$ & $\alpha$    & $\tfrac{3}{4} (1 + \ri)$ & $\alpha$    & $\tfrac{3}{4} (1 + \ri)$\\
 
 $D$   & $\bar \alpha$    &   $\tfrac{9}{8}$   &   $\alpha$  &$\tfrac{9}{8} \ri$ & $\alpha'$ &$\tfrac{10}{9} \ri$  & $\bar \alpha'$   &$\tfrac{10}{9}$\\    
 
 $D_{\sharp}$ & $\alpha'$   &$\tfrac{5}{6} (1 + \ri)$ & $\bar \alpha'$   &$\tfrac{5}{6} (1 - \ri)$ & $\bar \alpha$   &$\tfrac{5}{6} (1 - \ri)$ & $\alpha$   &$\tfrac{5}{6} (1 + \ri)$ \\
 
 $E$   & $\bar \alpha$    &$\tfrac{5}{4}$ & $\bar \alpha$    &$\tfrac{5}{4}$ & $\bar \alpha$    &$\tfrac{5}{4}$ & $\bar \alpha$    &$\tfrac{5}{4}$\\ 
 
 $F$    & $Z$    &$\tfrac{4}{3}$ & $Z$    &$\tfrac{4}{3}$ & $Z$    &$\tfrac{4}{3}$ & $Z$    &$\tfrac{4}{3}$ \\
 
 $F_{\sharp}$    & $\alpha$    &$(1+\ri)$ & $\alpha$    &$(1+\ri)$ & $\alpha$    &$(1+\ri)$ & $\alpha$    &$(1+\ri)$\\
 
 $G$   & $\bar \alpha$    &$\tfrac{3}{2}$  & $\bar \alpha$    &$\tfrac{3}{2}$ & $\bar \alpha$    &$\tfrac{3}{2}$  & $\bar \alpha$    &$\tfrac{3}{2}$ \\
 
 $G_{\sharp}$    &  $\alpha$    &$\tfrac{9}{8} (1 + \ri)$ & $\bar \alpha$ &$\tfrac{9}{8} (1 - \ri)$ &$\bar \alpha'$    &$\tfrac{10}{9} (1-\ri)$ &$\alpha'$    &$\tfrac{10}{9} (1+\ri)$\\ 
 
 $A$   & $\bar \alpha'$     &$\tfrac{5}{3}$ & $\alpha'$    &$\tfrac{5}{3}$ &$\alpha$ &$\tfrac{5}{3}$ &$\bar \alpha$ &$\tfrac{5}{3}$\\
 
 $A_{\sharp}$    & $\alpha$    &$\tfrac{5}{4} (1 + \ri)$ & $\alpha$    &$\tfrac{5}{4} (1 + \ri)$  & $\alpha$    &$\tfrac{5}{4} (1 + \ri)$ & $\alpha$    &$\tfrac{5}{4} (1 + \ri)$\\ 
 
 $B$    & $Z$    &$\tfrac{4}{3} (1 + \ri)$ & $Z$    &$\tfrac{4}{3} (1 + \ri)$ & $Z$    &$\tfrac{4}{3} (1 + \ri)$ & $Z$    &$\tfrac{4}{3} (1 + \ri)$\\ 
 
 $C'$ & $\bar \alpha$    &$2$ & $\bar \alpha$    &$2$  & $\bar \alpha$    &$2$ & $\bar \alpha$  &$2$  \\
  
  \hline
\end{tabular}
\caption{Four variants of the Complex System II scale based on three semitones. For each variant, the first column gives the aggregation sequence and the second column gives the values of the notes.  \vspace{0.25cm} \newline Notation: $\alpha = \tfrac{3}{4} (1 + \ri)$, \,$\alpha' = \tfrac{80}{81} \alpha$, \,$Z = \frac{16}{15}.$}

\end{table}

\newpage
\begin{table}
\vspace{3cm}
\begin{tabular}{ |p{.6cm}||p{.4cm}|p{1.5cm}||p{.4cm}|p{1.3cm}||}
 \hline
\multicolumn{5}{|c|}{\textbf{Complex System III}}\\
 \hline
Note & \multicolumn{2}{c||}{$X_2 = \rho, \,X_3 = Z$} &\multicolumn{2}{c||}{$X_2 = Z, \,X_3 = \rho$}\\
 \hline
 $C$ &1 &1 &1 & 1\\
 
 $C_{\sharp}$ & $\alpha$ & $\tfrac{3}{4}(1+\ri)$ & $\alpha$ & $\tfrac{3}{4}(1+\ri)$  \\
 
 $D$ & $\rho$ & $\tfrac{25}{32}(1+\ri)$ & $Z$ & $\tfrac{4}{5}(1+\ri)$ \\    
 
 $D_{\sharp}$ & $Z$ & $\tfrac{5}{6}(1+\ri)$ & $\rho$ & $\tfrac{5}{6}(1+\ri)$ \\
 
 $E$ & $\bar \alpha$ & $\tfrac{5}{4}$ & $\bar \alpha$ & $\tfrac{5}{4}$ \\ 
 
 $F$  & $Z$ & $\tfrac{4}{3}$ &  $Z$ & $\tfrac{4}{3}$\\
 
 $F_{\sharp}$  & $\alpha$ & $(1+\ri)$ & $\alpha$ & $(1+\ri)$\\
 
 $G$  & $\bar \alpha$ & $\tfrac{3}{2}$ & $\bar \alpha$ & $\tfrac{3}{2}$\\
 
 $G_{\sharp}$  & $\rho$ & $\frac{25}{16}$ & $Z$ & $\tfrac{8}{5}$ \\ 
 
 $A$  & $Z$ & $\tfrac{5}{3}$ &  $\rho$ & $\tfrac{5}{3}$\\
 
 $A_{\sharp}$  & $\alpha$ & $\tfrac{5}{4}(1+\ri)$ & $\alpha$ & $\tfrac{5}{4}(1+\ri)$ \\ 
 
 $B$  & $Z$ & $\tfrac{4}{3}(1+\ri)$ & $Z$ & $\tfrac{4}{3}(1+\ri)$\\ 
 
 $C'$ & $\bar \alpha$ & 2 & $\bar \alpha$ & 2\\
  
  \hline
\end{tabular}
\caption{Two variants of the Complex System III chromatic scale based on three semitones. For each scale, the first column gives the aggregation sequence and the second column gives the values of the notes of the scale. \vspace{0.25cm} \newline Notation: $\alpha = \tfrac{3}{4} (1 + \ri)$, \,$Z = \frac{16}{15}$, \,$\rho = \tfrac{25}{24}$.}
\end{table}

\newpage
\begin{table}
\vspace{3cm}
\begin{tabular}{ |p{.6cm}||p{.5cm}|p{1.9cm}||p{.4cm}|p{1.3cm}|| p{.4cm}|p{.8cm}|| p{.4cm}|p{.8cm}||}
 \hline
\multicolumn{9}{|c|}{\textbf{Real and Complex Pythagorean Scales based on Two Semitones}}\\
 \hline
Note & \multicolumn{2}{c||}{{\bf Complex Pyth}} &\multicolumn{2}{c||}{{\bf Shadow I}}&\multicolumn{2}{c||}{{\bf Shadow  II}}&\multicolumn{2}{c||}{{\bf Real Pyth}}\\
 \hline
 $C$   & 1    &1 &1 & 1&1&1&1&1\\
 
 $C_{\sharp}$   & $\alpha$    & $\tfrac{3}{4} (1 + \ri)$ &$\theta$&$\tfrac{3}{4}\sqrt{2}$ &$\theta$&$\tfrac{3}{4}\sqrt{2}$& $\sigma$    & $\tfrac{256}{243}$ \\
 
 $D$   & $\bar \alpha$    &   $\tfrac{9}{8}$ &$\theta $&$\tfrac{9}{8}$& $\theta $&$\tfrac{9}{8}$&$\tau$ &$\tfrac{9}{8}$\\    
 
 $D_{\sharp}$   & $\alpha$   &$\tfrac{27}{32} (1+\ri)$ &$\theta$ &$\tfrac{27}{32} \sqrt{2}$& $\theta$ &$\tfrac{27}{32} \sqrt{2}$&$\sigma$   &$\tfrac{32}{27}$ \\
 
 $E$   & $\bar \alpha, \sigma$    &$\tfrac{81}{64},\, \tfrac{8}{9}(1+\ri)$ &$\theta, \sigma$&$\tfrac{81}{64}, \,\tfrac{8}{9} \sqrt{2}$& $\nu$&$\tfrac{3}{4}\sqrt[4]{8}$&$\tau$  &$\tfrac{81}{64}$ \\ 
 
 $F$   & $\sigma, \bar\alpha$    &$\tfrac{4}{3}$  &$\sigma,  \theta$ &$\tfrac{4}{3}$& $\nu$  &$\tfrac{4}{3}$&$\sigma$    &$\tfrac{4}{3}$  \\
 
 $F_{\sharp}$   & $\alpha$    &$1+\ri$ &$\theta$&$\sqrt{2}$& $\theta$&$\sqrt{2}$&$\sigma$    &$\frac{1024}{729}$ \\
 
 $G$   & $\bar \alpha$    &$\tfrac{3}{2}$ &$\theta $&$\tfrac{3}{2}$& $\theta $&$\tfrac{3}{2}$&$\tau$    &$\tfrac{3}{2}$  \\
 
 $G_{\sharp}$   & $\alpha$    &$\tfrac{9}{8} (1 + \ri)$ &$\theta $&$\tfrac{9}{8} \sqrt{2}$& $\theta $&$\tfrac{9}{8} \sqrt{2}$& $\sigma$    &$\tfrac{128}{81}$  \\ 
 
 $A$   & $\bar \alpha$    &$\tfrac{27}{16}$ &$\theta $&$\tfrac{27}{16}$& $\theta $&$\tfrac{27}{16}$&$\tau$    &$\tfrac{27}{16}$  \\
 
 $A_{\sharp}$   & $\alpha, \sigma$  &$\tfrac{81}{64} (1+\ri),\, \tfrac{16}{9}$ &$\theta, \sigma$&$\tfrac{81}{64}\sqrt{2}, \, \tfrac{16}{9}$& $\nu$&$\tfrac{3}{2}\sqrt[4]{2}$ &$\sigma$    &$\tfrac{16}{9}$ \\ 
 
 $B$   & $\sigma, \alpha$    &$\tfrac{4}{3}(1 + \ri)$ &$\sigma, \theta $&$\tfrac{4}{3}\sqrt{2}$& $\nu $&$\tfrac{4}{3}\sqrt{2}$&$\tau$    &$\tfrac{243}{128}$ \\ 
 
 $C'$ & $\bar \alpha$    &$2$ &$\theta$&2&$\theta$&2& $\sigma$    &$2$  \\
  
  \hline
\end{tabular}
\caption{Chromatic scales generated by two semitones for (a) the complex Pythagorean scale, (b) the associated real shadow scales, and (c) a version of the traditional real Pythagorean scale. For each scale the first column gives the aggregation sequence and the second column gives the values of the notes of the scale. The variants that arise in each case are indicated. \vspace{0.25cm} \newline Notation: $\alpha = \tfrac{3}{4} (1 + \ri)$, \,$\theta  = \tfrac{3}{4}\sqrt{2}$, \,$\nu = \frac{8}{9} \sqrt[4]{2} $, \,$\sigma = \frac{256}{243}$, \,$\tau = \frac{2187}{2048}$.}
\end{table}

\newpage
\begin{table}
\begin{tabular}{|p{1.5cm}|p{9cm}|}
 \hline
\multicolumn{2}{|c|}{\textbf{List of Complex Intervals}}\\
\hline
$\pm 1$, $\pm i$ & unit (\,$\epsilon$\,) \\
$\frac{3+4i}{5}$ & five-limit phase factor (\,$\zeta$\,) \\  

$\frac{81}{80}$ & syntonic comma (\,$\kappa$\,) \\  

$\frac{25}{24}$ & minor chromatic semitone, small half tone (\,$\rho$\,) \\  
$\frac{20}{27} (1+i)$ & grave complex semitone (\,$\alpha'\,$) \\  
$\frac{256}{243}$ & minor Pythagorean semitone, diesis (\,$\sigma$\,) \\  
$\frac{1}{3} (3+i)$ & minor complex semitone  (\,$\beta$\,)\\  
$\frac{135}{128}$ & major chromatic semitone, major limma \\  
$\frac{3}{4} (1+i)$ &  major complex semitone $(\,\alpha$\,) \\  
$\frac{16}{15}$ & minor diatonic semitone  ($Z$) \\  
$\frac{2187}{2048}$ & major Pythagorean semitone, apotome  (\,$\tau$\,)\\  
$\frac{27}{25}$ & major diatonic semitone \\  

$\frac{10}{9}$ & minor whole tone ($Y$) \\  
$\frac{1}{2} (1+2i)$ &  minor complex whole tone (\,$\gamma$\,) \\  
$\frac{9}{8}$ & major whole tone ($X$) \\  
$\frac{4}{5} (1+i)$ &  major complex whole tone (\,$\delta$\,) \\  

$\frac{5}{6} (1+i)$ & complex minor third \\  
$\frac{32}{27}$ & minor Pythagorean  third \\  
$\frac{27}{32} (1+i)$ & complex minor Pythagorean  third \\  
$\frac{6}{5}$ & minor third \\  

$\frac{5}{4}$ & major third \\  
$\frac{81}{64}$ & major Pythagorean third \\  

$\frac{4}{3}$ & perfect fourth \\  
$\frac{27}{20}$ & acute fourth \\  
\hline

\end{tabular}
\caption{List of frequently occurring real and complex intervals, in order of magnitude.}
\end{table}

\newpage
\begin{table}
\begin{tabular}{|p{1.5cm}|p{9cm}|}
 \hline
\multicolumn{2}{|c|}{\textbf{List of Complex Intervals (continued)}}\\
\hline

$\frac{25}{18}$ & alternate tritone \\  
$\frac{1024}{729}$ & Pythagorean tritone \\  
$\frac{45}{32}$ & diatonic tritone \\  
$1+i$ & complex tritone (\,$\xi$\,)  \\  
$\frac{64}{45}$ & complementary diatonic tritone \\  
$\frac{36}{25}$ & complementary alternate tritone \\

$\frac{40}{27}$ & grave fifth \\  
$\frac{3}{2}$ & perfect fifth \\  
$\frac{25}{16}$ & augmented fifth \\  

$\frac{128}{81}$ & minor Pythagorean sixth \\  
$\frac{1}{2} (3-i)$ & complex minor sixth \\  
$\frac{9}{8} (1+i)$ & complex minor Pythagorean sixth \\  
$\frac{8}{5}$ & minor sixth \\  

$\frac{5}{3}$ & major sixth \\  
$\frac{27}{16}$ & major Pythagorean sixth \\  

$\frac{5}{4} (1+i)$ & complex minor seventh \\  
$\frac{16}{9}$ & minor Pythagorean seventh \\  
$\frac{81}{64} (1+i)$ & complex minor Pythagorean seventh \\  
$\frac{9}{5}$ & minor seventh \\  

$\frac{15}{8}$ & major seventh \\  
$\frac{4}{3} (1+i)$ & complex major seventh \\  
$\frac{243}{128}$ & major Pythagorean seventh \\  
$2$ & octave \\  
\hline

\end{tabular}
\caption{Frequently occurring real and complex intervals, in order of magnitude. Continued from previous page.}
\end{table}


\end{document}